\documentclass[11pt]{article}
\usepackage[utf8]{inputenc}
\usepackage{amsfonts,epsfig}
\usepackage[hyphens]{url}
\usepackage{hyperref}
\usepackage{breakurl}

\usepackage{relsize}
\usepackage{fancyvrb}
\usepackage{amssymb}
\usepackage{geometry} 
\usepackage{sectsty} 
\usepackage{caption} 
\usepackage{subcaption} 
\usepackage[normalem]{ulem} 
\sectionfont{\sffamily\bfseries\upshape\large}
\subsectionfont{\sffamily\bfseries\upshape\normalsize} 
\subsubsectionfont{\sffamily\mdseries\upshape\normalsize}
\makeatletter
\renewcommand\@seccntformat[1]{\csname the#1\endcsname.\quad}
\makeatother\renewcommand{\bibitem}{\vskip 2pt\par\hangindent\parindent\hskip-\parindent}

\makeatletter
\def\@maketitle{%
  \begin{center}%
  \let \footnote \thanks
    {\large \@title \par}%
    {\normalsize
      \begin{tabular}[t]{c}%
        \@author
      \end{tabular}\par}%
    {\small \@date}%
  \end{center}%
}
\makeatother

\title{\bf Beyond subjective and objective in statistics\footnote{We thank Sebastian Weber, Jay Kadane, Arthur Dempster, Michael Betancourt, Michael Zyphur, E. J. Wagenmakers, Deborah Mayo, James Berger, Prasanta Bandyopadhyay, Laurie Paul, Jan-Willem Romeijn, Gianluca Baio, Keith O'Rourke, and Laurie Davies for helpful comments.}\vspace{.1in}}
\author{Andrew Gelman\footnote{Department of Statistics and Department of Political Science, Columbia University, New York.} \and Christian Hennig\footnote{Department of Statistical Science, University College London.}}
\date{5 August 2015}
\begin{document}
\maketitle
\thispagestyle{empty}

\begin{abstract}
We argue that the words ``objectivity'' and ``subjectivity'' in statistics discourse are used in a mostly unhelpful way, and we propose to replace each of them with broader collections of attributes, with objectivity replaced by {\em transparency}, {\em consensus}, {\em impartiality}, and {\em correspondence to observable reality}, and subjectivity replaced by awareness of {\em multiple perspectives} and {\em context dependence}.  The advantage of these reformulations is that the replacement terms do not oppose each other.  Instead of debating over whether a given statistical method is subjective or objective (or normatively debating the relative merits of subjectivity and objectivity in statistical practice), we can recognize desirable attributes such as transparency and acknowledgment of multiple perspectives as complementary goals.  We demonstrate the implications of our proposal with recent applied examples from pharmacology, election polling, and socioeconomic stratification.
\end{abstract}

\section{Introduction}

\subsection{Motivation}
We can't do statistics without data, and as statisticians much of our efforts revolve around modeling the links between data and substantive constructs of interest.  We might analyze national survey data on purchasing decisions as a way of estimating consumers' responses to economic conditions; or gather blood samples over time on a sample of patients with the goal of estimating the metabolism of a drug, with the ultimate goal of coming up with a more effective dosing schedule; or we might be performing a more exploratory analysis, seeking clusters in a multivariate dataset with the aim of discovering patterns not apparent in simple averages of raw data.

As applied researchers we are continually reminded of the value of integrating new data into an analysis, and the balance between data quality and quantity.  In some settings it is possible to answer questions of interest using a single clean dataset, but more and more we are finding that this simple textbook approach does not work.

External information can come in many forms, including (a) recommendations on what variables to adjust for non-representativeness of a survey or imbalance in an experiment or observational study; (b) the extent to which outliers should be treated as regular, erroneous, or as indicating something that is meaningful but essentially different from the main body of observations; (c) substantial information on the role of variables, including potential issues with measurement, confounding, and substantially meaningful effect sizes; (d) population distributions that are used in poststratification, age adjustment, and other procedures that attempt to align inferences to a common population of interest; (e) restrictions such as smoothness or sparsity that serve to regularize estimates in high-dimensional settings; (f) the choice of functional form in a regression model (which in economics might be chosen to work with a particular utility function, or in public health might be motivated based on success in similar studies in the literature); and (g) numerical information about particular parameters in a model.  Of all these, only the final item is traditionally given the name ``prior information'' in a statistical analysis, but all can be useful in serious applied work. Other relevant information concerns not the data generating process but rather how the data and results of an analysis are to be used or interpreted.

We were motivated to write the present paper because we felt that our applied work, and that of others, was impeded because of the conventional framing of certain statistical analyses as subjective.  It seemed to us that, rather than being in opposition, subjectivity and objectivity both had virtues that were relevant in making decisions about statistical analyses.  We have earlier noted (Gelman and O'Rourke, 2015) that statisticians typically choose their procedures based on non-statistical criteria, and philosophical traditions and even the labels attached to particular concepts can affect real-world practice.

In this paper we reassess objectivity and subjectivity, exploding each into several sub-concepts, and we demonstrate the relevance of these ideas for three of our active applied research projects:  a hierarchical population model in pharmacology, a procedure for adjustment of opt-in surveys, and a cluster analysis of data on socioeconomic stratification.  We hope that readers will likewise see the relevance of these ideas in their own applied work, where decisions must be made about how to combine information of varying quality from different sources.

\subsection{Objectivity and subjectivity}

The continuing interest in and discussion of objectivity and subjectivity in statistics is, we believe, a necessary product of a fundamental tension in science:  On one hand, scientific claims should be impersonal in the sense that a scientific argument should be understandable by anyone with the necessary training, not just by the person promulgating it, and it should be possible for scientific claims to be evaluated and tested by outsiders.  On the other hand, the process of scientific inference and discovery involves individual choices; indeed, scientists and the general public celebrate the brilliance and inspiration of greats such as Einstein, Darwin, and the like, recognizing the roles of their personalities and individual experiences in shaping their theories and discoveries, and philosophers of science have studied the interplay between personal attitudes and scientific theories (Kuhn, 1962).  Thus it is clear that objective and subjective elements arise in the practice of science, and similar considerations hold in statistics.

Within statistics, though, discourse on objectivity and subjectivity is at an impasse.  Ideally these concepts would be part of a consideration of the role of different sorts of information and assumptions in statistical analysis, but instead they often seemed to be used in restrictive and misleading ways.

One problem is that the terms ``objective'' and ``subjective'' are loaded with so many associations and are often used in a mixed descriptive/normative way.  Scientists whose methods are branded  as subjective have the awkward choice of either saying, No, we are really objective, or else embracing the subjective label and turning it into a principle.  From the other direction, scientists who use methods labeled as objective often seem so intent on eliminating subjectivity from their analyses, that they end up censoring themselves.  This happens, for example, when researchers rely on $p$-values but refuse to recognize when their choice of analysis is contingent on data and that the theory behind the $p$-values is therefore invalidated (as discussed by Simmons, Nelson, and Simonsohn, 2011, and Gelman and Loken, 2014): significance testing is often used as a tool for a misguided ideology that leads researchers to hide, even from themselves, the iterative searching process by which a scientific theory is mapped into a statistical model or choice of data analysis (Box, 1983).  More generally, misguided concerns about subjectivity can lead researchers to avoid incorporating relevant and available information into their analyses and adapting the analyses appropriately to their research questions and potential uses of their results.

Many users of the terms ``objective'' and ``subjective'' in discussions concerning statistics do not acknowledge that these terms are quite controversial in the philosophy of science, and that they are used with a variety of different meanings and are therefore prone to misunderstandings. 

\section{Our proposal} \label{sprop}

We propose when talking about statistics to replace, where ever possible, the words ``objectivity'' and ``subjectivity'' with broader collections of attributes, with objectivity replaced by {\em transparency}, {\em consensus}, {\em impartiality}, and {\em correspondence to observable reality}, and subjectivity replaced by awareness of {\em multiple perspectives} and {\em context dependence}.

The advantage of this reformulation is that the replacement terms do not oppose each other.  Instead of debating over whether a given statistical method is subjective or objective (or normatively debating the relative merits of subjectivity and objectivity in statistical practice), we can recognize attributes such as transparency and acknowledgment of multiple perspectives as complementary goals.

\subsection{``Transparency,'' ``consensus,''  ``impartiality,''  and ``correspondence to observable reality,'' instead of ``objectivity''}
 
Merriam-Webster defines ``objective'' as ``based on facts rather than feelings or opinions: not influenced by feelings'' and ``existing outside of the mind: existing in the real world'' (actually the concept is quite controversial, see Section \ref{sobjconc}).
Science is practiced by human beings, who only have access to the real world through interpretation of their perceptions. Taking objectivity seriously as an ideal, scientists need to make the sharing of their perceptions and interpretations possible. When 
applied to statistics, the implication is that the choices in the data analysis (including the prior distribution, if any, but also the model for the data, methodology, and the choice of what information to include in the first place) should be motivated based on factual, externally verifiable information and transparent criteria.  This is similar to the idea of the concept of ``institutional decision analysis'' (Section 9.5 of Gelman, Carlin, et al., 2013), under which the mathematics of formal decision theory can be used to ensure that decisions can be justified based on clearly-stated criteria.  Different stakeholders will disagree on decision criteria, and different scientists will differ on statistical modeling decisions, so, in general, there is no unique ``objective'' analysis, but we can aim at communicating and justifying analyses in ways that support scrutiny and eventually consensus.  Similar thoughts have motivated the slogan ``transparency is the new objectivity'' in journalism (Weinberger, 2009).

In the context of statistical analysis, a key aspect of objectivity is therefore a process of {\em transparency}, in which the choices involved are justified based on external, potentially  verifiable sources or at least transparent considerations (ideally accompanied by sensitivity analyses if such considerations leave alternative options open), a sort of ``paper trail'' leading from external information, through modeling assumptions and decisions about statistical analysis, all the way to inferences and decision recommendations. But transparency is not enough. We hold that science aims at {\em consensus} in potentially free exchange (see Section \ref{sourpersp} for elaboration), which is one reason that the current crisis of non-replication is taken so seriously in psychology (Yong, 2012).  Transparency contributes to this building of consensus by allowing scholars to trace the sources and information used in statistical reasoning (Gelman and Basb\o ll, 2013). Furthermore, scientific consensus, as far as it deserves to be called ``objective,'' requires rationales, clear arguments and motivation, and elucidation how this relates to already existing knowledge. Following generally accepted rules and procedures counters the dependence of results on the personalities of the individual researchers, although there is always a danger that such generally accepted rules and procedures are inappropriate or suboptimal for the specific situation at hand. In any case, consensus can only be achieved if researchers attempt to be {\em impartial} by taking into account competing perspectives, avoiding to favor pre-chosen hypotheses, and being open to criticism. 

The world outside the observer's mind plays a key role in usual concepts of objectivity. Finding out about the real world is seen by many as the major objective of science, and this suggests correspondence to reality as the ultimate source of scientific consensus. This idea is not without its problems and meets some philosophical opposition; see Section \ref{sobjconc}. We acknowledge that the ``real world'' is only accessible to human beings through observation, and that scientific observation and measurement cannot be independent of human preconceptions and theories.
As statisticians we are concerned with making general statements based on systematized observations, and this makes {\em correspondence to observed reality} a core concern regarding objectivity. 
This is not meant to imply 
that empirical statements about observations
are the only meaningful ones that can be made about reality; we think that 
scientific theories that cannot be verified (but potentially be falsified) 
by observations 
are meaningful thought constructs, particularly because observations are 
truly independent of thought constructs.

Formal statistical methods
contribute to objectivity as far as they contribute to the fulfillment of these desiderata, particularly by making procedures and their implied rationales transparent and unambiguous.

For example, Bayesian statistics is commonly characterized as ``subjective'' by Bayesians and non-Bayesians alike. But depending on how exactly prior distributions are interpreted and used (see Sections \ref{ssubjbayes}--\ref{sfalbayes}), they fulfill or aid some or all of the virtues listed above.  Priors make the researchers' prior point of view transparent; different approaches of interpreting them provide different rationales for consensus; ``objective Bayesians'' (see Section \ref{sobjbayes}) try to make them impartial; and if suitably interpreted (see Section \ref{sfalbayes}) they can be properly grounded in observations.

\subsection{``Multiple perspectives'' and ``context dependence,'' instead of ``subjectivity''} \label{smulcosub}

Merriam-Webster defines ``subjective'' as ``relating to the way a person experiences things in his or her own mind'' and ``based on feelings or opinions rather than facts.'' Science is normally seen as striving for objectivity, and therefore acknowledging subjectivity is not popular in science. But as noted above already, reality and the facts are only accessible through individual personal experiences. Different people bring different information and different viewpoints to the table, and they will use scientific results in different ways. In order to enable clear communication and consensus, differing perspectives need to be acknowledged, which contributes to transparency and thus to objectivity. Therefore, subjectivity is important to the scientific process. Subjectivity is valuable in statistics in that it represents a way to incorporate the information coming from differing perspectives. 

We propose to replace the concept of ``subjectivity'' with {\em awareness of  multiple perspectives} and {\em context dependence}.  To the extent that subjectivity in statistics is a good thing, it is because information truly is dispersed, and, for any particular problem, different stakeholders have different goals. A counterproductive implication of the idea that science should be ``objective'' is that there is a tendency in the communication of statistical analyses to either avoid or hide decisions that cannot be made in an automatic, seemingly ``objective'' fashion by the available data.  Given that all observations of reality depend on the perspective of an observer, interpreting science as striving for a unique (``objective'') perspective is illusory. Multiple perspectives are a reality to be reckoned with and should not be hidden. Furthermore, by avoiding personal decisions, researchers often waste opportunities to adapt their analyses appropriately to the context, the specific background and their specific research aims, and to communicate their perspective more clearly. Therefore we see the acknowledgment of multiple perspectives and context dependence as virtues, making clearer in which sense subjectivity can be productive and helpful.  

The term ``subjective'' is often used to characterize aspects of certain statistical procedures that cannot be derived in an automatic manner from the data to be analyzed, such as Bayesian prior distributions and tuning parameters (for example, the proportion of trimmed observations in trimmed means, or the threshold in wavelet smoothing). Such decisions are entry points for multiple perspectives and context dependence. The first decisions of this kind are typically the choice of data to be analyzed and the family of statistical models to be fit.

To connect with the other half of our proposal, the recognition of different perspectives should be done in a transparent way.  We should not say we set a tuning parameter to 2.5 (say) just because that is our belief. Rather, we should justify the choice explaining clearly how it supports the research aims. This could be by embedding the choice in a statistical model that can ultimately be linked back to observable reality and empirical data, or by reference to desirable characteristics (or avoidance of undesirable artifacts) of the methodology given the use of the chosen parameter; actually, many tuning parameters are related to such characteristics and aims of the analysis rather than to some assumed underlying 
``belief'' (see Section \ref{stransformation}). In some cases, such a justification may be imprecise, for example because background knowledge may be only qualitative and not quantitative or not precise enough to tell possible alternative choices apart, but often it can be argued that even then conscious tuning or specification of a prior distribution comes with benefits compared to using default methods of which the main attraction often is that seemingly ``subjective'' decisions can be avoided.

To consider an important example, regularization requires such decisions. Default priors on regression coefficients are used to express the belief that coefficients are typically close to zero, and from a non-Bayesian perspective, lasso shrinkage can be interpreted as encoding an external assumption of sparsity. Sparsity assumptions can be connected to an implicit or explicit model in which problems are in some sense being sampled from some distribution or probability measure of possible situations; see Section \ref{sfalbayes}. This general perspective (which can be seen as Bayesian with an implicit prior on states of nature, or classical with an implicit reference set for the evaluation of statistical procedures) provides a potential basis to connect choices to experience; at least it makes transparent what kind of view of reality is encoded in the choices.

Tibshirani (2014) writes that enforcing sparsity is not primarily motivated by beliefs about the world, but rather by benefits such as computability and interpretability, hinting at the fact that considerations other than being ``close to the real world'' often play an important role in statistics and more generally in science. Even in areas such as social science where no underlying truly sparse structure exists, imposing sparsity can have advantages such as supporting stability (Gelman, 2013). 

In a wider sense, if one is performing a linear or logistic regression, for example, and considering options of maximum likelihood, lasso, or hierarchical Bayes with a particular structure of priors, all of these choices are ``subjective'' in the sense of encoding aims regarding possible outputs and assumptions, and all are ``objective'' as far as these aims and assumptions are made transparent and the assumptions can be justified based on past data and ultimately be checked given enough future data.  So the conventional labeling of Bayesian analyses or regularized estimates as ``subjective'' misses the point.

Alternatively to basing it on past data, the choice of tuning parameter can be based on knowledge of the impact of the choice on results and a clear explanation why a certain impact is desired or not. In  robust statistics, for example, the breakdown point of some methods can be tuned and may be chosen lower than the optimal 50\%, because if there is a too large percentage of data deviating strongly from the majority, one may rather want the method to deliver a compromise between all observations, but if the percentage of outliers is quite low, one may rather want them to be disregarded, with borderline percentages depending on the application (particularly on to what extent outliers are interpreted as erroneous observations rather than as somewhat special but still relevant cases). 



\subsection{A list of specific objective and subjective virtues} \label{svirtue}
To summarize the above discussion, virtues that are often referred to as ``objective'' 
include:
\begin{enumerate}
\item Transparency:
  \begin{enumerate}
  \item Clear and unambiguous definitions of concepts,
  \item Open planning and following agreed protocols,
  \item Full communication of reasoning, procedures, and potential limitations;
  \end{enumerate}
\item Consensus:
  \begin{enumerate}
  \item Accounting for relevant knowledge and existing related work,
  \item Following generally accepted rules where possible and reasonable,
  \item Provision of rationales for consensus and unification;
  \end{enumerate}
\item Impartiality:
  \begin{enumerate}
  \item Thorough consideration of relevant and potentially  
competing theories and points of view,
  \item Thorough consideration and if possible removal 
of potential biases: factors that may 
jeopardize consensus and the intended interpretation of results,
  \item Openness to criticism and exchange;
  \end{enumerate}
\item Correspondence to observable reality:
  \begin{enumerate}
  \item Clear connection of concepts and models to observables,
  \item Clear conditions for reproduction, testing, and falsification.
  \end{enumerate}
\end{enumerate}
This last bit is a challenge in statistics, as reproduction, testing, and falsification can only be assessed probabilistically in any real, finite-sample setting.


What about subjectivity?
The term ``subjective'' is often used as opposite to ``objective'' and as such
often meant to be opposed to scientific virtues, or to be something
that cannot fully be avoided and that therefore has to be only 
grudgingly accepted.

But subjective perspectives are the building blocks
for scientific consensus, and therefore there are also scientific virtues 
associated with subjectivity:
\begin{enumerate}
\item Awareness of multiple perspectives,
\item Awareness of context dependence:
\begin{enumerate}
\item Recognition of dependence on specific contexts and aims,
\item Honest acknowledgment of the researcher's position, goals, 
experiences, and subjective point of view.
\end{enumerate}
\end{enumerate}

In the subsequent discussion we shall label the items in the above lists as O1a--O4b or O1--O4 for groups 
of items (``O'' for ``connected to
objectivity''), and S1, S2 (S2a, S2b) for the items connected to subjectivity.
Our intention is to sketch a system of virtues that allows a
more precise and detailed discussion where issues of objectivity and
subjectivity are at stake. 

We are aware that in some situations some of these
virtues may oppose each other, for example 
``consensus'' can contradict ``awareness of
multiple perspectives,'' and indeed dissent is essential to scientific progress.  This tension between impersonal consensus and creative debate is an unavoidable aspect of science. 
Sometimes the consensus can only be that there are different
legitimate points of view. Furthermore, the listed virtues are not 
all fully autonomous; clear reference to observations may be both 
a main rationale for consensus and a key contribution to transparency; 
and the three subjective virtues contribute to both transparency and
openness to criticism and exchange.

Not all items on the list
apply to all situations. For example, 
in the following section we will apply the list to
the foundations of statistics, but the items O1c
and S2b rather apply to specific studies.

\section{Applied examples} \label{sexa}

In conventional statistics, assumptions are commonly minimized.  Classical statistics and econometrics is often framed in terms of robustness, with the goal being methods that work with minimal assumptions.  But the decisions about what information to include and how to frame the model---these are typically buried, not stated formally as assumptions but just baldly stated:  ``Here is the analysis we did \dots,'' sometimes with the statement or implication that these have a theoretical basis but typically with little clear connection between subject-matter theory and details of measurements.  From the other perspective, Bayesian analyses are often boldly assumption-based but with the implication that these assumptions, being subjective, need no justification and cannot be checked from data.

We would like statistical practice, Bayesian and otherwise, to move toward more transparency regarding the steps linking theory and data to models, and recognition of multiple perspectives in the information that is included in this paper trail and this model. In this section we show how we are trying to move in this direction in some of our recent research projects.  We present these examples not as any sort of ideals but rather to demonstrate how we are grappling with these ideas and, in particular, the ways in which active awareness of the concepts of transparency, consensus, impartiality, correspondence to
observable reality, multiple perspectives and context dependence
is changing our applied work.

\subsection{A hierarchical Bayesian model in pharmacology}

Statistical inference in pharmacokinetics/pharmacodynamics involves many challenges:  data are indirect and often noisy; the mathematical models are nonlinear and computationally expensive, requiring the solution of differential equations; and parameters vary by person but often with only a small amount of data on each experimental subject.  Hierarchical models and Bayesian inference are often used to get a handle on the many levels of variation and uncertainty (see, for example, Sheiner, 1984, and Gelman, Bois, and Jiang, 1996).

One of us is currently working on a project in drug development involving a Bayesian model that was difficult to fit, even when using advanced statistical algorithms and software.  Following the so-called folk theorem of statistical computing (Gelman, 2008), we suspected that the problems with computing could be attributed to a problem with our statistical model.  In this case, the issue did not seem to be lack of fit, or a missing interaction, or unmodeled measurement error---problems we had seen in other settings of this sort.  Rather, the fit appeared to be insufficiently constrained, with the Bayesian fitting algorithm being stuck going through remote regions of parameter space that corresponded to implausible or unphysical parameter values.

In short, the model as written was only weakly identified, and the given data and priors were consistent with all sorts of parameter values that did not make scientific sense.  Our iterative Bayesian computation had poor convergence---that is, the algorithm was having difficulty approximating the posterior distribution---and the simulations were going through zones of parameter space that were not consistent with the scientific understanding of our pharmacology colleagues.

To put it another way, our research team had access to prior information that had not been included in the model.  So we took the time to specify a more informative prior.  The initial model thus played the role of a placeholder or default which could be elaborated as needed, following the iterative prescription of falsificationist Bayesianism (Box, 1980, Gelman et al., 2013, Section \ref{sfalbayes}).

In our experience, informative priors are not so common in applied Bayesian inference, and when they are used, they often seem to be presented without clear justification.  In this instance, though, we decided to follow the principle of transparency and write a note explaining the genesis of each prior distribution.  To give a sense of what we're talking about, we present a subset of these notes here:

\begin{small}
\begin{itemize}
\item $\gamma_1$: mean of population distribution of $\log
  (\mbox{BVA}_j^{\rm latent}/50)$, centered at 0 because the mean of
  the BVA values in the population should indeed be near 50.  We set
  the prior sd to 0.2 which is close to $\log(60/50)=0.18$ to indicate
  that we're pretty sure the mean is between 40 and 60.
\item $\gamma_2$: mean of pop dist of $\log (k^{\rm in}_{j} / k^{\rm
    out}_{j})$, centered at 3.7 because we started with $-2.1$ for
  $k^{\rm in}$ and $-5.9$ for $k^{\rm out}$, specified from the literature about the disease.  We use a sd of 0.5 to
  represent a certain amount of ignorance: we're saying that our prior
  guess for the population mean of $k^{\rm in} / k^{\rm out}$ could
  easily be off by a factor of $\exp(0.5)=1.6$.
\item $\gamma_3$: mean of pop dist of log $k^{\rm out}_{j}$, centered
  at $-5.8$ with a sd of 0.8, which is the prior that we were given
  before, from the time scale of the natural disease progression.
\item $\gamma_4$: $\log E^0_{\rm max}$, centered at 0 with sd
  2.0 because that's what we were given earlier.
\end{itemize}
\end{small}

We see this sort of painfully honest justification as a template for future Bayesian data analyses.  The above snippet certainly does not represent an exemplar of best practices, but we see it as a ``good enough'' effort that presents our modeling decisions in the context in which they were made.

To label this prior specification as ``objective'' or ``subjective'' would miss the point.  Rather, we see it as having some of the virtues of objectivity and subjectivity---notably, transparency (O1) and some aspects of consensus (O2) and awareness of multiple perspectives (S1)---while recognizing its clear imperfections and incompleteness.  Other desirable features would derive from other aspects of the statistical analysis---for example, we use external validation to approach correspondence to observable reality (O4), and our awareness of context dependence (S2) comes from the placement of our analysis within the larger goal, which is to model dosing options for a particular drug.

One concern about our analysis which we have not yet thoroughly addressed is sensitivity to model assumptions.  We have established that the prior distribution makes a difference but it is possible that different reasonable priors yield posteriors with greatly differing real-world implications, which would raise concern about consensus (O2) and impartiality (O3).  Our response to such concerns, if this sensitivity is indeed a problem, would be to more carefully document our choice of prior, thus doubling down on the principle of transparency (O1) and to compare to other possible prior distributions supported by other information, thus supporting impartiality (O3) and awareness of multiple perspectives (S1).

As with ``institutional decision analysis'' (Gelman et al., 2003, section 22.5), the point is not that our particular choices of prior distributions are ``correct'' (whatever that means), or optimal, or even good, but rather that they are transparent, and in a transparent way connected to knowledge. Subsequent researchers---whether supportive, critical, or neutral regarding our methods and substantive findings---should be able to interpret our priors (and, by implication, our posterior inferences) as the result of some systematic process, a process open enough that it can be criticized and improved as appropriate.

\subsection{Adjustments for pre-election polls}

Wang et al.\ (2014) describe another of our recent applied Bayesian research projects, in this case a statistical analysis that allows highly stable estimates of public opinion by adjustment of data from non-random samples.  The particular example used was an analysis of data from an opt-in survey conducted on the Microsoft Xbox video game platform, a technique that allowed the research team to, effectively, interview respondents in their living rooms, without ever needing to call or enter their houses.

The Xbox survey was performed during the two months before the 2012 U.S. presidential election.  In addition to offering the potential practical benefits of performing a national survey using inexpensive data, this particular project made use of its large sample size and panel structure (repeated responses on many thousands of Americans) to learn something new about U.S. politics:  we found that certain swings in the polls, which had been generally interpreted as representing large swings in public opinion, actually could be attributed to differential nonresponse, with Democrats and Republicans in turn being more or less likely to respond during periods where there was good or bad news about their candidate.  This finding was consistent with some of the literature in political science (see Erikson, Panagopoulos, and Wlezien, 2004), but the Xbox study represented an important empirical confirmation.

Having established the potential importance of the work, we next consider its controversial aspects.  For many decades, the gold standard in public opinion research has been probability sampling, in which the people being surveyed are selected at random from a list or lists (for example, selecting households at random from a list of addresses or telephone numbers and then selecting a person within each sampled household from a list of the adults who live there).  From this standpoint, opt-in sampling of the sort employed in the Xbox survey lacks a theoretical foundation, and the estimates and standard errors thus obtained (and which we reported in our research papers) do not have a clear statistical interpretation.

This criticism---that inferences from opt-in surveys lack a theoretical foundation--is interesting to us here because it is {\em not} framed in terms of objectivity or subjectivity.  We do use Bayesian methods for our survey adjustment but the criticism from certain survey practitioners is not about adjustment but rather about the data collection:  they take the position that no good adjustment is possible for data collected from a non-probability sample.

As a practical matter, our response to this criticism is that nonresponse rates in national random-digit-dialed telephone polls are currently in the range of 90\%, which implies that real-world surveys of this sort are essentially opt-in samples in any case:  If there is no theoretical justification for non-random samples then we are all dead, which leaves us all with the choice to either abandon statistical inference entirely when dealing with survey data, or to accept that our inferences are model-based and do our best (Gelman, 2014c).

We shall now express this discussion using the criteria from Section \ref{svirtue}.  Probability sampling has the clear advantage of transparency (O1) in that the population and sampling mechanism can be clearly defined and accessible to outsiders, in a way that an opt-in survey such as the Xbox is not.  In addition, the probability sampling has the benefits of consensus (O2), at least in the United States, where such surveys have a long history and are accepted in marketing and opinion research.  Impartiality (O3) and correspondence to observable reality (O4) are less clearly present because of the concern with nonresponse, just noted.  We would argue that the large sample size and repeated measurements of the Xbox data, coupled with our sophisticated hierarchical Bayesian adjustment scheme, put us well on the road to impartiality (through the use of multiple sources of information, including past election outcomes, used to correct for biases in the form of known differences between sample and observation) and correspondence to observable reality (in that the method can be used to estimate population quantities that could be validated from other sources).

Regarding the virtues associated with subjectivity, the various adjustment schemes represent awareness of context dependence (S2) in that the choice of variables to match in the population depend on the context of political polling, both in the sense of which aspects of the population are particularly relevant for this purpose, and in respecting the awareness of survey practitioners of what variables are predictive of nonresponse. The researcher's subjective point of view is involved in the choice of exactly what information to include in weighting adjustments and exactly what statistical model to fit in regression-based adjustment. Users of probability sampling on grounds of ``objectivity'' may shrink from using such judgments, and may therefore ignore valuable information from the context.


\subsection{Transformation of variables in cluster analysis for socioeconomic stratification} \label{stransformation}

Cluster analysis
aims at grouping together similar objects and separating dissimilar ones, and as such is based, explicitly or implicitly, on some measure of dissimilarity measure. Defining such a measure, for example using some set of variables characterizing the objects to be clustered, can
involve many decisions. Here we consider an example of Hennig and Liao (2013), where we clustered data from the 2007
U.S. Consumer Finances Survey, comprising variables on income, savings, housing,
education, occupation, number of checking and savings accounts, and life 
insurance with the aim of data-based exploration of socioeconomic stratification. The choice of variables and the decisions of how they are selected, transformed, 
standardized, and weighted has a strong impact on the results of the cluster 
analysis. This impact depends to some extent on the  
clustering technique that is afterward applied to the resulting 
dissimilarities, but will typically be considerable, even for cluster analysis
techniques that are not directly based on dissimilarities. 
One of the various issues
discussed by Hennig and Liao (2013) was the transformation of the variables 
treated as continuous (namely income and savings amount), with the view of
basing a cluster analysis on a Euclidean distance after transformation, 
standardization, and weighting of variables. 

There is some literature on choosing transformations, but the 
usual aims of transformation, namely achieving approximate additivity, linearity, equal variances, or normality, are often not relevant for cluster 
analysis, where such assumptions only apply to model-based
clustering, and only within the clusters, which are not known before
transformation.  

The rationale for transformation when setting up a dissimilarity measure
for clustering is of a different kind. The measure needs to
formalize appropriately which objects are to be treated as ``similar'' or
``dissimilar'' by the clustering methods, and should therefore be 
put into the same or different clusters, respectively. In other words, the
formal dissimilarity between objects should match what could be called the 
``interpretative dissimilarity'' between objects. This is an
issue involving subject-matter knowledge that cannot be decided by the data alone. 

Hennig and Liao (2013) argue that the interpretative dissimilarity between
different savings amounts is governed rather by ratios than by differences, so that
\$2 million of savings is seen as about as dissimilar from 
\$1 million, as \$2,000 is dissimilar from \$1,000. This
implies a logarithmic transformation. We do not argue that
there is a precise argument that privileges the log
transformation over other transformations that achieve something similar, and
one might argue from intuition that even taking logs may not be strong enough.
We therefore recognize that any choice of transformation is a provisional
device and only an approximation to an ideal ``interpretative dissimilarity,''
even if such an ideal exists.

In the dataset, there are no negative savings values as there is no information on debts, but there are many people who report zero savings, and it is conventional to kluge the
logarithmic
transformation to become $x \mapsto \log(x+c)$ with some $c>0$.
Hennig and Liao then point out that, in this example, the choice of $c$ has a 
considerable impact on clustering. The number of people with very small but 
nonzero savings in the dataset is rather small.
Setting  $c=1$, for example, the transformation creates a substantial
gap between the zero savings group and people with fairly low (but not very 
small) amounts of savings, and of course this choice is also sensitive to scaling (for example, savings might be coded in dollars, or in thousands of dollars).  The subsequent cluster analysis (done by 
``partitioning around medoids''; Kaufman and Rousseeuw, 1990)  would therefore
separate the zero savings group strictly; no person with zero savings would
appear together in a cluster with a person with nonzero savings. For larger 
values for $c$, the dissimilarity between the zero savings group 
and people with a low savings amount becomes effectively small enough that
people with zero savings could appear in clusters together with other people,
as long as values on other variables are similar enough. 

We do not believe that there is a true value of $c$. Rather, 
clusterings arising from different choices of $c$ are legitimate but imply
different interpretations. The clustering for $c=1$ is based on treating 
the zero savings group as very special, whereas the clustering for $c=200$,
say, implies that a difference in savings between 0 and \$100 is taken
as not such a big deal (although it is a bigger deal in any case than 
the difference
between \$100 and \$200). Similar considerations hold for issues such as
selecting and weighting variables and coding ordinal variables. 
 
It can be frustrating to the novice in cluster analysis that such decisions
for which there do not seem to be an objective basis can make such a
difference, and there is apparently a very strong temptation to ignore the 
issue and to just choose $c=1$, which may look ``natural'' in the sense that it
maps zero onto zero, or even to avoid transformation at all in order to avoid the
discussion, so that no obvious lack of objectivity strikes the reader.
Having the aim of socioeconomic stratification in mind, though, 
it is easy to argue 
that clusterings that result from ignoring the issue are less desirable and
useful than a clustering obtained from making a however imprecisely grounded
decision choosing a $c>1$, therefore avoiding either separation of the zero
savings group as a clustering artifact or
an undue domination of the clustering by people with large savings
in case of not applying 
any transformation at all. 

We believe that this kind of tuning problem that cannot be
interpreted as estimating an unknown true constant (and does therefore
not lend itself naturally to an approach through a Bayesian prior)
is not exclusive to cluster analysis, and is often hidden in 
presentations of data analyses.

In Hennig and Liao (2013), we pointed out the issue and did some sensitivity analysis 
about the strength of the impact of the choice of $c$ (O1, transparency).
The way we picked the $c$ in that paper made clear reference
to the context dependence, while being honest that the subject-matter knowledge in this case provided only weak guidelines for making this decision (S2).
We were also clear that alternative choices would amount to 
alternative perspectives rather than being just wrong (S1, O3).

The issue how to foster consensus and to make a connection to observable reality
(O2, O4) is of interest, but not treated here.


It is, however, problematic to establish rationales for consensus that are 
based on ignoring the context and potentially multiple perspectives. 
There is a tendency in the cluster analysis literature to seek
formal arguments for making such decisions automatically (see, for example,
Everitt et al., 2011, Section 3.7, on variable weighting; it is hard to find 
anything systematic in the clustering literature on transformations), 
for example trying to optimize ``clusterability'' of the dataset, or to prefer
methods that are less sensitive to such decisions, because this amounts
to making the decisions implicitly without giving the researchers access to 
them. In other words, the data are given the authority to determine not only 
which objects are similar (which is what we want them to do), but also what 
similarity should mean. The latter should be left to the researcher, although 
we acknowledge that the data can have a certain impact: for example the idea
that dissimilarity of savings amounts is governed by ratios rather than 
differences is connected to (but not determined by) 
the fact that the distribution of savings amounts
is skewed, with large savings amounts sparsely distributed.

\subsection{Testing for homogeneity against clustering}

Another feature of the cluster analysis in Hennig and Liao (2013) was a 
parametric bootstrap test for homogeneity against clustering, see also Hennig
and Lin (2015) for a more general elaboration. Clusterings can
be computed regardless of whether the data are clustered in a sense that is relevant for the application of interest. In this example, the test involved the construction of
a null model that captured the features of the dataset such as 
the dependence between variables and marginal distributions of the categorical 
variables as well as possible, without involving anything that could be
interpreted as clustering structure. As opposed to the
categorical variables, the marginal distributions 
of the ``continuous'' variables such as the transformed savings amount 
were treated as potentially indicating clustering, and therefore the null 
model used unimodal distributions for them. As test statistic we used a 
cluster validity statistic of the clustering computed on the data, with a
parametric bootstrap used to compute a clustering in the same manner
on data generated from the null model.

We used a classical significance test
rather than a Bayesian approach here because we were not interested in
posterior probabilities for either the null model to be true or prediction
of future observations. Rather the question of interest was whether the observed
clustering structure in the data (as measured by the validity index) could be
explained by a model without any feature that would be interpreted as ``real
clustering,'' regardless of whether or to what extent 
we believe this model or not. However, we deviated from classical 
significance test logic in some ways, particularly not using a test statistic that was
optimal test against any specific alternative, instead choosing a statistic
pointing in a rough direction (namely ``clustering'') from the null model.
Furthermore, setting up the null model required decisions on which potential
characteristics of the dataset would be interpretable as ``clustering,'' on 
could therefore not be incorporated in the null model that was to be interpreted
as ``non-clustering.'' A non-significant outcome of the test can then clearly 
be interpreted as no evidence in the data for real clustering, whereas
the interpretation of a significant outcome depends on whether we can argue
convincingly that the null model is as good as it gets at trying to model
the data without clustering structure. Setting up a straw man null model
for homogeneity and rejecting it would have been easy and not informative.

There is no point in arguing that our significance test was more objective
than for example a Bayesian analysis would have been, and actually our 
approach involved decisions such as the distinction between data 
characteristics interpreted as ``clustering'' or ``non-clustering'' and the
choice of a test statistic that were made by by considerations other than 
seemingly objective mathematical optimality or estimation from the data. 
Still the ultimate aim was to see whether the idea of a real 
clustering would be supported by the data (O4), in an impartial and transparent manner (O1, O3), trying hard to give the null model a fair chance to fit
the data, but involving context dependent judgment (S2) and the transparent
choice of a specific perspective (the chosen validity index) among a potential variety (S1), because we were after more qualitative statements than degrees of belief in certain models. 

\section{Objectivity and subjectivity in statistics and science}
\label{sobjsubj}
\subsection{Discussions within statistics}

In discussions of the foundations of statistics, objectivity and subjectivity are seen as opposites. Objectivity is typically seen as a good thing; many see it as a major requirement for good science. Bayesian statistics is often presented as being subjective because of the choice of a prior distribution. Some Bayesians (notably Jaynes, 2003, and Berger, 2006) have advocated an objective approach, whereas others (notably de Finetti, 1974) have embraced subjectivity. It has been argued that the subjective/objective distinction is meaningless because all statistical methods, Bayesian or otherwise, require subjective choices, but the choice of prior distribution is sometimes held to be particularly subjective because, unlike the data model, it cannot be determined for sure even in the asymptotic limit.  In practice, subjective prior distributions often have well known empirical problems such as overconfidence (Alpert and Raiffa, 1984, Erev, Wallsten, and Budescu, 1994), which motivates efforts to check and calibrate Bayesian models (Rubin, 1984, Little, 2012) and to situate Bayesian inference within an error-statistical philosophy (Mayo, 1996, Gelman and Shalizi, 2013).

De Finetti can be credited with acknowledging honestly that subjective decisions cannot be avoided in statistics, but it is misleading to think that the required subjectivity always takes the form of prior belief.  The confusion arises from two directions:  first, prior distributions are not necessarily any more subjective than other aspects of a statistical model; indeed, in many applications priors can and are estimated from data frequencies (see Chapter 1 of Gelman, Carlin, et al., 2013, for several examples).  Second, somewhat arbitrary choices come into many aspects of statistical models, Bayesian and otherwise, and therefore we think it is a mistake to consider the prior distribution as the exclusive gate at which subjectivity enters a statistical procedure.

The objectivity vs.\ subjectivity issue also arises with statistical methods that require tuning parameters; decision boundaries such as the significance level of tests; and decisions regarding inclusion, exclusion, and transformation of data in preparation for analysis.

On one hand, statistics is sometimes said to be the science of defaults:  most applications of statistics are performed by non-statisticians who adapt existing general methods to their particular problems, and much of the research within the field of statistics involves devising, evaluating, and improving such generally applicable procedures (Gelman, 2014b).  It is then seen as desirable that any required data-analytic decisions or tuning are performed in an objective manner, either determined somehow from the data or justified by some kind of optimality argument.

On the other hand, practitioners must apply their subjective judgment in the choice of what method to use, what assumptions to invoke, and what data to include in their analyses.  Even using ``no need for tuning'' as a criterion for method selection or prioritizing bias, for example, or mean squared error, is a subjective decision. Settings that appear completely mechanical involve choice:  for example, if a researcher has a checklist saying to apply linear regression for continuous data, logistic regression for binary data, and Poisson regression for count data, he or she still has the option to code a response as continuous or to use a threshold to define a binary classification.  And such choices can be far from trivial; for example, when modeling elections or sports outcomes, one can simply predict the winner or instead predict the numerical point differential or vote margin.  Modeling the binary outcome can be simpler to explain but in general will throw away information, and subjective judgment arises in deciding what to do in this sort of problem (Gelman, 2013a). And in both classical and Bayesian statistics, subjective choices arise in defining the sample space and considering what information to condition on.

\subsection{Discussions in other fields}

Scholars in humanistic studies such as history and literary criticism have considered the ways in which differently-situated observers can give different interpretations to what Luc Sante calls the ``factory of facts.''  In political arguments, controversies often arise over ``cherry picking'' or selective use of data, a concern we can map directly to the statistical principle of random or representative sampling, and the more general idea that information used in data collection be included in any statistical analysis (Rubin, 1978).  In a different way, the concepts of transference and counter-transference, central to psychoanalysis, live at the boundary of personal impressions and measurable facts, all subject to the constraint that, as Philip K. Dick put it, ``Reality is that which, when you stop believing in it, doesn't go away.''

The social sciences have seen endless arguments over the relative importance of objective conditions and what Keynes (1936) called ``animal spirits.''  In macroeconomics, for example, the debate has been between the monetarists who tend to characterize recessions as necessary consequences of underlying economic conditions (as measured, for example, by current account balances, business investment, and productivity), and the Keynesians who focus on more subjective factors such as stock market bubbles and firms' investment decisions.  These disagreements also turn methodological, with much dispute, for example, over the virtues and defects of various attempts to objectively measure the supply and velocity of money, or consumer confidence, or various other inputs to economic models.  The interplay between objective and subjective effects also arises in political science, for example in the question of whether to attribute the political successes of a Ronald Reagan or a Bill Clinton to their charisma and appealing personalities, to their political negotiating skills, or simply to periods of economic prosperity that would have made a success out of just about any political leader.  Again, these disputes link to controversies regarding research methods:  a focus on objective, measurable factors can be narrow, but with a more subjective analysis it can be difficult to attain a scientific consensus.  In fields such as social work it has been argued that one must work with subjective realities in order to make objective progress (Saari, 2005), but this view is relevant to science more generally.

In the social and physical sciences alike (as well as in hybrid fields such as psychophysics), the twentieth century saw an intertwining of objectivity and subjectivity.  From one direction, Heisenberg's uncertainty principle told us that, at the quantum level, measurement depends fundamentally on the observation process, an insight that is implicit in modern statistics and econometrics with likelihood functions, measurement-error models, and sampling and missing-data mechanisms being manifestations of observation models.  So in that sense there is no pure objectivity.  From the other direction, psychologists have continued their effort to scientifically measure personality traits and subjective states.  For example, Kahneman (1999) defines ``objective happiness'' as ``the average of utility over a period of time.''  Whether or not this definition makes much sense, it illustrates a movement in the social and behavioral sciences to measure, in supposedly objective manners, what might previously have been considered unmeasurable.

Much of these discussions are relevant to statistics because of the role of quantification. There is an ideology widespread in many areas of science that sees quantification and numbers and their statistical analysis as key tools for objectivity. An important function of quantitative scientific measurement is the production of observations that are thought of as independent of individual points of view. Apart from the generally difficult issue of measurement validity, the focus on what can be quantified, however, may narrow down what can be observed, and may not necessarily do the measured entities justice, see the examples from political science and psychology above. Another example is the use of quantitative indicators for human rights in different countries; although it has been argued that it is of major importance that such indicators should be objective to have appropriate impact on political decision making (Candler et al., 2011), many aspects of their definition and methodology are subject to controversy and reflect specific political interests and views (Merry, 2011), and we think that it will help the debate to communicate such indicators transparently together with their limitations and the involved decisions rather than to sell them as objective and unquestionable. In many places the present paper may read as if we treat the observations to be analyzed by the statisticians as given, but we acknowledge the central importance of measurement and the benefits and drawbacks of quantification. See Porter (1996), Desrosieres (2002), Douglas (2009) for more discussion of the connection between quantification and objectivity. As with choices in statistical modeling and analysis, we believe that when considering measurement the objective/subjective antagonism is less helpful than a more detailed discussion of what quantification can achieve and what its limitations are.

\subsection{Concepts of objectivity} \label{sobjconc}
Discussions involving objectivity and subjectivity often suffer from objectivity having multiple meanings, in statistics and elsewhere (much of the following discussion will focus on the term ``objectivity''; subjectivity is often considered as the opposite of objectivity and as such implicitly defined). Ambiguity in these terms is often ignored. We believe that such discussions can become clearer by referring to the meanings that are relevant in any specific situation instead of using the ambiguous terms ``objectivity'' and ``subjectivity'' without further explanation. 

Lorraine Daston has explored the ways in which objectivity has been used as a way to generalize scientific inquiry and make it more persuasive.  As Daston (1992) puts it, scientific objectivity ``is conceptually and historically distinct from the ontological aspect of objectivity that pursues the ultimate structure of reality, and from the mechanical aspect of objectivity that forbids interpretation in reporting and picturing scientific results.''  The core of the current use of the term ``objectivity'' is the idea of impersonality of scientific statements and procedures. According to Daston and Galison (2007), the term has only been used in this way in science from the mid-nineteenth century; before then, ``objective'' and ``subjective'' were used with meanings almost opposite from the current ones and did not play a strong role in discussions about science.  Daston (1994) specifically addresses changing concepts of subjectivity and objectivity of probabilities, and Zabell (2011) traces the historical development of these concepts. 

The idea of independence of the individual subject can be applied in various 
ways. Megill (1994) listed four basic senses of objectivity: ``absolute
objectivity'' in the sense of ``representing the things as they really are''
(independently of an observer), ``disciplinary objectivity'' referring to a 
consensus among experts within a discipline and highlighting the role of 
communication and negotiation, ``procedural objectivity'' in the
sense of following rules that are independent of the individual researcher,
and ``dialectical 
objectivity.'' The latter somewhat surprisingly involves subjective 
contributions, because it refers to active human ``objectification'' required to
make phenomena communicable and measurable so that they can then be treated 
in an objective way so that different subjects can understand them in
the same way. Statistics for example relies on the construction of well
delimited populations and categories within which averages and probabilities 
can be defined; see Desrosieres (2002). 

Daston and Galison (2007) call the ideal of 
scientific images that attempt to capture reality in an unmanipulated way 
``mechanical objectivity'' as opposed to ``structural objectivity,'' which
emerged from the insight of scientists and philosophers such as Helmholtz and
Poincare that observation of reality cannot exclude the observer
and will never be as reliable and pure as ``mechanical objectivists'' would 
hope. Instead, ``structural objectivity'' 
refers to mathematical and logical structures. Porter (1996)
lists the ideal of impartiality of observers as another sense of objectivity,
and highlights the important role of quantitative and formal reasoning for
concepts of objectivity because of their potential for removing ambiguities.
In broad agreement with interpretations already listed (and covered by our
virtues), Reiss and Sprenger (2014) group key aspects of objectivity 
into the categories
``faithfulness to facts,'' ``absence of normative commitments and 
value-freedom,'' and ``absence of personal bias.'' Fuchs (1997) notes that 
various modern meanings of objectivity rather refer to the absence of 
subjectivity and all kinds of biasing factors than to something positive.

To us, the most problematic aspect of the term ``objectivity'' is that it incorporates normative and descriptive aspects, and that these are often not clearly delimited. For example, a statistical method that does not require the specification of any tuning parameters is objective in a descriptive sense (it does not require decisions by the individual scientist). Often this is presented as an advantage of the method without further discussion, implying objectivity as a norm, but depending on the specific situation the lack of flexibility caused by the impossibility of tuning may actually be a disadvantage (and indeed can lead to subjectivity at a different point in the analysis, when the analyst must make the decision of whether to use an auto-tuned approach in a setting where its inferences do not appear to make sense).
The frequentist interpretation of probability is objective in the sense that it locates probabilities in an objective world that exists independently of the observer, but the definition of these probabilities requires a subjective definition of a reference set. Although some proponents of frequentism consider its objectivity (in the sense of impersonality, conditional on the definition of the reference set) as a virtue, this property is ultimately only descriptive; it does not imply on its own that such probabilities indeed exist in the objective world, nor that they are a worthwhile target for scientific inquiry. 

The interpretation of objectivity as a scientific virtue is connected to what 
are seen to be the aims and values of science. 
Scientific realists hold that 
finding out the truth about the observer-independent reality is the 
major aim of science. This makes
``absolute objectivity'' as discussed above a core scientific ideal, as which 
it is still popular. But
observer-independent reality is only accessible through human observations,
and the realist ideal of objectivity has been branded as metaphysical,
meaningless, and illusory by positivists including Karl Pearson (1911), 
and more contemporarily by 
empiricists such as van Fraassen (1980). In the latter groups, 
objectivity is seen
as a virtue as well, although for them it does not refer to 
observer-independent reality but rather to a standardized, disciplined, and 
impartial application of scientific methodology enabling academic consensus 
about observations.

Reference to observations is an element that the 
empiricist, positivist, and realist ideas of objectivity have in common;
Mayo and Spanos (2010)  see checking theories against experience by means 
of what they call ``error statistics'' as a central tool to ensure objectivity,
which according to them is concerned with finding out about reality in an 
unbiased manner\footnote{Mayo emphasizes that her approach does not require 
being a realist; according to our reading, she is in any case concerned with 
observer-independent reality, as opposed to the positivists, 
without subscribing to naive and all too 
optimistic ideas about what we can know about it.}. In contrast, van Fraassen 
(1980) takes observability and the ability of theory to account for 
observed facts as objective from an {\em anti}-realist perspective. His 
construal of observability depends on the context, theory, and means of
observation, and his concept of objectivity is conditional on these 
conditions of observation, assuming that at least acceptance of 
observations and observability given these conditions should not depend on the 
subject.

Daston and Galison (2007) portray the rise of ``mechanical objectivity'' as a scientific virtue in reaction to shortcomings of the earlier scientific ideal of ``truth-to-nature,'' which  refers to the idea that science should discover and present an underlying ideal and universal (Platonic) truth below the observed phenomena. The move towards mechanical objectivity, inspired by the development of photographic techniques, implied a shift of perspective; instead of producing pure and ideal ``true'' types the focus moved to capturing nature ``as it is,'' with all irregularities and variations that had been suppressed by a science devoted to ``truth-to-nature.'' Increasing insight in the shortcomings and the theory-dependence of supposedly objective observational techniques led to the virtue of ``trained judgment'' as a response to mechanical objectivity. According to Daston and Galison (2007), the later virtues did not simply replace the older ones, but rather supplemented them, so that nowadays all three still exist in science. 


Another typology of objectivity was set up by Douglas (2004), who distinguishes three modes of objectivity, namely human interaction with the world (connected to our ``correspondence to observable reality''), individual thought processes (connected to our ``impartiality'') and processes to reach an agreement (connected to our ``consensus'' and ``transparency''). These modes are subdivided into different ``senses.'' Regarding human interaction with the world, Douglas distinguishes objectivity connected to human manipulation and intervention and objectivity connected to stability of results when taking multiple approaches of observation. Regarding individual thought processes, a interesting distinction is made between prohibition against using values in place of evidence and against using any values at all. Douglas suspects that the latter is hard to achieve and will rather encourage sweeping issues under the carpet. She writes that ``hiding the decisions that scientists make, and the important role values should play in those decisions, does not exclude values.'' A third sense is the conscious attempt to be value-neutral. The three proposed senses of objectivity regarding processes to reach an agreement are the use of generally agreed procedures, exploration of whether and to what extent consensus exists, and an interactive discursive attempt to achieve consensus.

Further distinctions regarding objectivity appear in the 
philosophical literature. Reiss and Sprenger (2014) distinguish the objectivity of a process, such as inference or procedure, from the objectivity of an outcome. Some of our aspects of objectivity, such as impartiality, concern the former; while others, such as correspondence to observable reality, concern the latter; but the connection is not always clear. Following Reichenbach (1938), there is much discussion in the philosophy of science concerning the distinction between the ``context of discovery'' and the ``context of justification,'' with some arguing that subjective impact is problematic for the latter but not the former. Some, however, challenge the idea that these two contexts can be appropriately separated and that one can avoid the impact of subjective values on justification, see Reiss and Sprenger (2014) for an overview.

Objectivity has also been criticized on the grounds that, as attractive 
as it may seem
as an ideal, it is illusory. This criticism has to refer to a specific 
interpretation of objectivity, and a weaker interpretation of objectivity may
still seem to critics to be a good thing: van Fraassen agrees with Kuhn 
(1962) and
others that ``absolute objectivity'' is an illusion and that access to reality
is dependent of the observer, but he still holds that objectivity conditional on
a system of reference is a virtue. But there is even criticism of the idea that
objectivity, possible or not, 
is desirable. From a particular feminist point of view, MacKinnon (1987) 
wrote: ``To look at the world objectively is to objectify it.'' Striving 
for objectivity itself is seen here as a specific and potentially harmful
perspective, implying a denial of the specific conditions of an observer's 
point of view. A similar point was made by Feyerabend (1978). Maturana (1988)
critically
discussed the ``explanatory path of objectivity-without-parenthesis'' in which 
observers deny personal responsibility for their positions based on a supposedly
privileged access to an objective reality; he 
appreciated a more perspective-dependent attitude, which he called 
``objectivity-with-parenthesis.''

Fuchs (1997) gives an overview of controversies regarding the role of 
objectivity in communication and discourse, looking at ideas such as
that objectivity
may be merely a rhetorical device, or a tool of power, or, on the other hand, 
a tool to defend suppressed views against power. Although he is critical of 
such ideas if interpreted in a reductionist manner, in his own theory he
portrays objectivity as a communicative ``medium'' of science, which is 
to some extent constitutive and essential for science, particularly for 
fostering consensus, although it comes with its own ``blind spot.'' 
We can connect with such a sociological perspective 
regarding the aspect that our criticism of objectivity rather points at its
role in the (statistics) discourse, which we do not see as indispensable, 
than at its meanings.

A recent paper titled ``Let's not talk about objectivity''  (Hacking, 2015) suggests that statistics is not alone in handling the concepts of objectivity and subjectivity in a messy way, and that we are not alone in advocating that such discourse should be replaced by looking at more specific virtues of scientific work.

However, despite our proposal to replace these terms in statistics, we appreciate the deep insight manifest in the philosophical controversies about objectivity and subjectivity and in the existing attempts to define them in a meaningful way. The reader will realize that our proposal was strongly influenced by the elaboration of different senses and types of objectivity in the philosophy of science, as discussed above.
We also respect the view that it is valuable to stick to a concept of objectivity that can serve to distinguish valid from (in some way) biased inference and science from pseudo-science. Deborah Mayo even suggested to us that the term ``objective'' could refer to an appropriate use of all our positive attributes combined, whereas what we are really (and correctly) opposed to is some kind of misleading ``fake'' notion of objectivity. However, facing some of the damage done by more common and much less thought through uses of ``objectivity'' and ``subjectivity'' in everyday exchanges, we still believe that our proposal will result in an improvement of the discussion culture in statistics overall.

\subsection{Our attitude toward objectivity and subjectivity in science}
\label{sourpersp}
The attitude taken in the present paper is based on Hennig (2010), which was in turn inspired by constructivist philosophy 
(Maturana, 1988, von Glasersfeld, 1995) and
distinguishes personal reality, social reality, and 
observer-independent reality. According to this perspective, 
human inquiry starts from observations that are
made by personal observers (personal reality). Through communication, 
people share observations and generate social realities that go beyond a
personal point of view.  These shared realities include for example measurement procedures 
that standardize observations, and mathematical models that connect 
observations to an abstract formal system that is meant to create a thought 
system cleaned from individually different point of views. 
Nevertheless, human beings only 
have access to observer-independent reality through personal observations and 
how these are brought together in social reality. 

According to Hennig (2010), science aims at arriving at a view of 
reality that is stable and reliable and can be agreed freely by general 
observers and is therefore
as observer-independent as possible. In this sense we see  
objectivity
as a scientific ideal. But at the same time we acknowledge what gave rise to
the criticism of objectivity: the existence of different 
individual perspectives and also of perspectives that  differ between 
social systems, and therefore the ultimate inaccessibility of a reality 
that is truly independent of observers, is a basic human condition. Objectivity
can only be attributed by observers, and if observers disagree about what
is objective, there is no privileged position from which this can be decided.
Ideal objectivity can never be achieved.

This does not imply, however, that scientific disputes can never be resolved by scientific means.  Yes, there is an element of ``politics'' involved in the adjudication of scholarly disagreements, but, as we shall discuss, the norm of {\em transparency} and other norms associated with both objectivity and subjectivity can advance such discussions.  In general no particular observer has a privileged position but this does not mean that all positions are equal.  We recognize subjectivity not to throw up our hands and give up on the possibility of scientific consensus but as a first step to exploring and, ideally, reconciling, the multiple perspectives that are inevitable in nearly any human inquiry.

Denying the existence of different legitimate subjective perspectives and
of their potential to contribute to scientific inquiry cannot make sense 
in the name of objectivity. Heterogeneous points of view cannot 
be dealt with by imposing authority. Our attitude values the attempt to 
reach scientific agreement between different perspectives, but ideally 
such an agreement is reached by free exchange between the different points 
of view. In practice, however, agreement will not normally be universal, and
in order to progress, science has to aim at a more restricted agreement 
between experts who have enough background knowledge to either make sure that 
the agreement about something new is in line with what was already 
established earlier, or to know that and how it requires a revision of
existing knowledge. But the resulting agreement is still intended to be 
potentially open for everyone to join or to challenge.  
Therefore, in science there is always a tension between the ideal of
general agreement and the reality of heterogeneous perspectives.

Furthermore our attitude to science is based on the idea that consensus is possible regarding stable and reliable statements about the observed reality
(which may require elaborate measurement procedures),
and that science aims at nontrivial knowledge in the sense that it
makes statements about observable reality that can and should 
be checked and potentially
falsified by observation.

Although there is no objective access 
to observer-independent reality, we acknowledge that there is an almost
universal human experience of a reality perceived as located outside the 
observer and as not controllable by the observer. This reality is a target
of science, although it cannot be taken for granted that it is indeed
independent of the observer. We are therefore ``active
scientific realists'' in the sense of Chang (2012), who writes: ``I take reality as
whatever is not subject to one's will, and knowledge as an ability
to act without being frustrated by resistance from reality. This
perspective allows an optimistic rendition of the pessimistic induction,
which celebrates the fact that we can be successful in
science without even knowing the truth. The standard realist
argument from success to truth is shown to be ill-defined and
flawed.'' This form of realism
is not in contradiction to the criticism of realism
by van Fraassen or the arguments against the desirability of certain forms of
objectivity by constructivists or feminists as outlined above.
Active scientific realism implies that finding out the 
truth about objective reality is not the ultimate aim of science, but that 
science
rather aims at 
supporting human actions. This means that scientific methodology has to 
be assessed relative to the specific aims and actions connected to its use.
Another irreducible subjective element in science, apart from multiple 
perspectives on reality, is therefore the aim of scientific inquiry, which
cannot be standardized in an objective way. A typical statistical instance
of this is how much prediction accuracy in a restricted setting is valued 
compared with parsimony and interpretability.
   
Because science aims at agreement, communication is central to science, as are
transparency and 
techniques for supporting the clarity of communication. Among these techniques
are formal and mathematical language, standardized measurement procedures,
and scientific models.
Objectivity as we see it is therefore a scientific ideal that can never 
fully be
achieved. As much as science aims for objectivity, it has to acknowledge
that it can only be built from a variety of subjective perspectives
through communication.

\section{Decomposing subjectivity and objectivity in the foundations of statistics} \label{sfound}

In this section, we use the above list of virtues to revisit aspects
of the discussion on fundamental approaches to statistics, for which 
the terms ``subjective'' and ``objective'' typically play a dominant role.
We discuss
what we perceive to be the major streams of the foundations of statistics, but
within each of these streams there exist several different approaches, which we
cannot cover completely in such a paper; rather we sketch the
streams somewhat roughly and refer to only a single or a few leading authors
for details where needed.

Here, we distinguish between interpretations of probability, and approaches for
statistical inference. Thus, we take frequentism to be an interpretation of 
probability, which does not necessarily imply that Fisherian or Neyman-Pearson
tests are preferred to Bayesian methods, despite the fact that frequentism 
is more often associated with the former than with the latter. 

We shall go through several philosophies of statistical inference, for each laying out the connections we see to the virtues of objectivity and subjectivity outlined in Section \ref{svirtue}.

Exercising awareness of multiple perspectives, we emphasize that we do not 
believe that one of these philosophies is the correct or best one, nor
do we claim that reducing the different approaches to a single one would
be desirable. What is lacking here is not unification, but rather, often, 
transparency about which interpretation of probabilistic outcomes is 
intended when applying statistical modeling to specific problems. 
Particularly, we think that, depending on the situation, 
both ``aleatory'' or ``epistemic''
approaches to modeling uncertainty are legitimate and worthwhile, 
referring to data generating processes in observer-independent reality
on one hand and rational degrees of belief on the other.

\subsection{Frequentism} \label{sfreq}
We label  ``frequentism'' as the identification of the probability of an 
event in a certain experiment with a limiting relative
frequency of occurrences if
the experiment were to be carried out infinitely often in some kind of 
independent manner. Frequentist statistics is based on evaluating procedures based on a long-term average over a ``reference set'' of  hypothetical replicated data sets.
%
In the wider sense, we call  probabilities  ``frequentist'' when they
formalize observer-independent 
tendencies or propensities of experiments to yield certain outcomes 
(see, for example, Gillies, 2000), which are thought of as replicable and 
yielding a behavior under infinite replication as suggested by what is 
assumed to be the ``true'' probability model. 

The frequentist mindset locates probabilities in 
the observer-independent world, so they are
in this sense objective. This objectivity, however, is model-based, as an 
infinite amount 
of actual replicates cannot exist, and most researchers, in most 
settings, would be
skeptical about truly identical replicates
and true independence or, when it comes
to observational studies, about whether observations can be interpreted as drawn
in a purely random manner from an appropriate reference set. 

The decision to adopt the frequentist interpretation of probability regarding
a certain phenomenon therefore requires idealization. It cannot be justified in
a fully objective way, which here means, referring to our list of virtues,
that it can neither be 
enforced by observation, nor is there general enough consensus that this
interpretation applies to any specific setup, although it is well discussed and
supported in some physical settings such as radioactive decay (O2, O4).
Once a frequentist model is adopted, however, it makes predictions about 
observations that can be checked, so the reference to the observable reality
(O4) is clear. 

There is some disagreement about whether the frequentist definition of 
probability is  
clear and unambiguous (O1a). 
On one hand, the idea of a tendency of an experiment to produce certain 
outcomes as manifested in observed and expected 
relative frequencies seems clear enough, given that the 
circumstances of the experiment are well defined and regardless of 
whether frequencies indeed
behave in the implied way. On the other hand, von Mises (1957) was not completely successful in his attempt to avoid involving 
stochastic independence and identity in the definition of frequentist 
probabilities 
through the concepts of the collective and the axiom of invariance
under place selection rules (Fine, 1973), and the issue has never been completely resolved. 

Frequentism implies that, in the observer-independent 
reality, true probabilities are unique, but there is considerable room
for multiple perspectives (S1) regarding the definition of replicable
experiments, collectives, or reference sets. The idea of replication
is often constructed in a rather creative way. For example, frequentist
time series models are used for time series data, implying an underlying
true distribution for every single time point, but there is no
way to repeat observations independently at the same time point. This actually
means that the effective sample size for time series data would be 1, if
replication was not implicitly constructed in the statistical model, for example by assuming 
independent innovations in ARMA-type models. Such models, or, more precisely, 
certain aspects of such models, can be checked against
the data, but even if such a check does not fail, it is still clear
that there is no such thing in observable 
reality, even approximately, as a marginal 
``true'' frequentist distribution of the value of the time series $x_t$ at 
fixed $t$, as implied by the model, because $x_t$ is strictly not replicable. 

The issue that useful statistical models require a construction of 
replication (or exchangeability) on some level by the statistician, 
is, as we discuss below, not confined to 
frequentist models. In order to provide a rationale for the essential statistical task of pooling
information from many observations to make inference relevant for future
observations, all these observations 
need to be assumed to somehow represent the same process.  

The appropriateness of such assumptions in a specific situation can often only
be tested in a quite limited way by observations. All kinds of informal
arguments can apply about why it is a good or bad idea to consider a
certain set of observations (or unobservable implied entities such as 
error terms and latent variables) as independent and identically distributed
frequentist replicates.

Unfortunately, although such an openness to multiple perspectives  and 
potential context-de\-pen\-dence (S2a) can be seen as positive from our 
perspective, these issues involved in the choices of a frequentist reference set are often  not clearly communicated and discussed.
The existence of a true model with implied reference set is typically taken for 
granted by frequentists, motivated at least in part by the desire for objectivity.

From the perspective taken here and
in Hennig (2010), the frequentist interpretation of probability can be 
adopted as an idealized model, a thought construction, 
without having to believe that frequentist probabilities really exist in the
observer-independent world (many criticisms of frequentism such as most of 
the issues raised in Hajek, 2009, refer to a belief in the ``existence'' of
limits of hypothetical sequences that are impossible in the real world). 
This can be justified, on a case-by-case basis,
if it is seen as useful for the scientific aims in the given situation,
for example because a specific frequentist model communicates (more or less)
clearly the scientist's view of a certain phenomenon (O1a), and implies the 
means for testing this against observations (O4).

\subsection{Error statistics} \label{serr}
The term ``error statistics'' was coined by the philosopher Deborah Mayo 
(1996). We use it here to refer to an approach to statistical inference
that is based on a frequentist interpretation of probability and methods
that can be characterized and evaluated by error probabilities.  
Traditionally these would be  
the Type I and Type II errors of Neyman-Pearson hypothesis testing, but the error-statistical perspective could also apply to other constructs such as errors of sign and magnitude (``Type S'' and ``Type M'' errors; Gelman and Carlin, 2014).
Mayo (1996) introduced another key concept for error statistics, ``severity,'' which is
connected with, but not identical to, the power of tests. It
serves to quantify the extent to which a test result can corroborate a
hypothesis (keeping in mind that testing specific statistical hypothesis 
can only ever shed light on specific aspects of a scientific theory of 
interest; and that a specific test can only corroborate a specific 
aspect of a hypothesized statistical model). The severity principle states that
a test result can only be evidence of the absence of a certain discrepancy 
from a (null) hypothesis, if the probability is high, given that such a 
discrepancy indeed existed, that the test result would have been less in line
with the hypothesis than what was observed\footnote{Mayo applies the term 
``severity'' also more generally, not confined to statistics.}.

According to Mayo and Spanos (2010), objectivity is a core concern of 
error statistics, which is specifically driven by providing methodology 
for reproduction, testing, and falsification (O4b). 
Mayo (2014) defined objective scientific measurement as being 
``relevant,'' ``reliably capable,'' and ``able to learn from error,'' 
which can be interpreted as the error-statistical rationale for consensus (O2c).
Error statistical methodology is portrayed as ``reliably capable'' as far as
its potential to produce inferential errors can be analyzed, and as far as the
resulting error probabilities are low. The ``ability to learn from error''
refers to erroneous hypotheses, rejected by an error statistical
procedure that optimally can pinpoint the reason for rejection and thus lead to
an improvement of the hypothesis, rather than errors of the inferential method.
The underlying idea, with which we  agree, 
is that learning from error is a main driving force
in science, a lifetime contract between the mode of statistical 
investigation and its object. This corresponds to Chang's active scientific 
realism mentioned above, 
and it implies that for Mayo the reference to observations
is central for objectivity.   

Mayo's ``relevance'' concerns the problem of inquiry of 
interest and is therefore related to virtue S2a, which we classified as
related to subjectivity. As Mayo attempts to defend the objectivity
of the error statistical approach against charges of subjectivity, she may not
be happy about this classification, but we agree with her that this is an
important virtue nonetheless, which, however, is not specifically connected
to error statistics.

The error probability characteristics of error statistical methods 
rely, in general, on model assumptions. In principle, these model assumptions can be
tested in an error statistical manner, too, and are therefore, according to
Mayo, no threat to the objectivity of the account. But this comes with two
problems. Firstly, derivations of statistical inference based on error probabilities typically
assume the model as fixed and do not account for prior model selection based
on the data. This issue has recently attracted some research (for example, 
Berk et al., 2013), but this still requires a transparent listing of all 
the possible modeling decisions that could be made (virtue O1b), which often
is missing, and which may not even be desirable as long as the methods are
used in an exploratory fashion (Gelman and Loken, 2014). Secondly, 
any dataset can be consistent with
many models, which can lead to divergent inferences. Davies (2014) illustrates 
this  with the analysis of a dataset on amounts of copper in drinking 
water, which can be fitted well by a Gaussian, a double 
exponential, and a comb distribution, but yields vastly
different confidence intervals for the center of symmetry (which is assumed 
to be the target of inference) under these three 
models\footnote{Davies (2014) uses the example for a wider discussion of 
modeling issues including regularization and 
defects of the likelihood.}.

Davies suggests that it is misleading to hypothesize models
or parameters to be ``true,'' and that one should instead take into 
account all 
models that are ``adequate'' for approximating the data in the sense that 
they are not rejected by tests based on features of the data the 
statistician is interested in, which does not require reference to 
unobservable true frequentist probabilities, but takes into account 
error probabilities as well. Such an approach is tied to the
observations in a more direct way without making metaphysical assumptions
about unobservable features of observer-independent reality (O1a, O4). 
However, it is possible that
such a metaphysical assumption is implicitly still needed if the 
researcher wants to use ``data approximating models'' to learn about 
observer-independent reality, and that the class of all adequate models
is too rich for meaningful inference 
(as in more standard frequentist treatments, 
Davies focuses on models with independent and identically distributed random variables or error terms).
Earlier work on robust statistics
(see Huber and Ronchetti, 2009) already introduced the idea of 
sets of models that neighbor a nominal model, from which the models in the 
neighborhood could not be reliably distinguished based on the data.

Even further flexibility in error statistical analyses comes from the fact
that the assumption of a single true underlying distribution does not 
determine the parametric or nonparametric family of distributions, within 
which the true distribution is embedded. Although Neyman and Pearson 
derived optimal tests considering specific alternatives to the null 
hypothesis, many kinds of alternatives and test statistics could be of potential
interest. Davies (2014) explicitly mentions the 
dependence of the choice of statistics for checking the adequacy of models
on the context and the researcher's aims (S2a) instead of relying on
Neyman-Pearson type optimality results.

Overall, there is no shortage of entry points for multiple
perspectives (S1) in the error statistical approach. This could be seen as 
something positive, but it runs counter to some extent to the way the
approach is advertised as objective by some of its proponents. 
Many frequentist and error statistical analyses could in our opinion
benefit from acknowledging honestly their flexibility and the researcher's
choices made, many of which cannot be determined by data alone.

\subsection{Subjectivist Bayesianism} \label{ssubjbayes}
We call ``subjectivist epistemic''  the interpretation of probabilities
as quantifications of strengths of belief of an individual, where probabilities
can be interpreted as derived from, or implementable through, bets that are
coherent in that no opponent can cause sure losses by setting up some
combinations of bets.  From this requirement of coherence, the usual probability axioms follow (O2c). 
Allowing conditional 
bets implies Bayes's theorem, and therefore, as far as inference 
concerns learning from observations about not (yet) observed hypotheses,
Bayesian methodology is used for subjectivist epistemic probabilities, hence 
the term ``subjectivist Bayesianism.'' 

A major proponent of subjectivist Bayesianism was Bruno de Finetti (1974). 
De Finetti was not against objectivity in general. He viewed observed
facts as objective, as well as mathematics and logic and certain formal
conditions of random experiments such as the set of possible outcomes. 
But he viewed uncertainty as something subjective and he held that
objective (frequentist) probabilities do not exist. He claimed that
his subjectivist Bayesianism appropriately takes into account both the 
objective (see above) and subjective (opinions about unknown facts based
on known evidence) components for probability evaluation. 
Given the degree of
idealization required for frequentism as discussed in Section \ref{sfreq},
this is certainly a legitimate position. 

In de Finetti's work the term ``prior'' refers to all probability assignments
using information external to the data at hand, with no fundamental distinction between the  
``parameter prior'' assigned to parameters in a model, and the form of the ``sampling distribution'' given a fixed parameter, in contrast to common Bayesian practice today,
in which the term ``prior'' is used to refer only to the parameter prior.   In the following discussion we shall use the term ``priors'' in de Finetti's general sense.

Regarding the list of virtues in Section \ref{svirtue}, de Finetti provides
a clear definition of probability (O1a) based on principles that 
he sought to establish as generally acceptable (O2c).  As opposed to
objectivist Bayesians, subjectivist Bayesians do not attempt to enforce 
agreement regarding prior distributions, not even given the same
evidence; still, de Finetti (1974) and other subjectivist Bayesians 
proposed rational principles for assigning prior probabilities. The difference
between the objectivist and subjectivist Bayesian point of view is rooted in
the general tension in science explained above; the subjectivist approach can
be criticized for not supporting agreement enough---conclusions based
on one prior may be seen as irrelevant for somebody who holds another one (O2c)---but can be defended for honestly acknowledging that 
prior information often does
not come in ways that allow a unique formalization (S2b). 
In any case it is vital 
that subjectivist Bayesians explain transparently how they arrive at their
priors, so that other researchers can decide to what extent they can
support the conclusions (O1c).  Such transparency is desirable in any statistical approach but is particularly relevant for subjective Bayesian models which cannot be rejected within the subjectivist paradigm 
in case of disagreement with observations.

In de Finetti's conception, probability assessments, prior and posterior,
can ultimately only concern observable events, because bets can only be 
evaluated if the experiment on which a bet is placed has an observable outcome,
and so there is a clear connection to observables (O3a). 

However, priors in the subjectivist Bayesian conception are not open to
falsification (O3b), because by definition they have to be fixed before 
observation. Adjusting the prior after having observed the data to be
analyzed violates coherence. The Bayesian system as derived from
axioms such as coherence (as well as those used by objectivist Bayesians; see
Section \ref{sobjbayes}) is designed to cover all aspects of learning from
data, including model selection and rejection, but this requires that all
potential later decisions are already incorporated in the prior, which 
itself is not interpreted as a testable 
statement about yet unknown observations. 
In particular this means that once a subjectivist Bayesian 
has assessed a setup as exchangeable a priori,  he or she cannot drop this assumption
later, whatever the data are (think of observing twenty zeroes, then twenty 
ones, then ten further zeroes in a binary experiment). This is a 
major problem, because
subjectivist Bayesians use de Finetti's theorem to justify working with
parameter priors and sampling models under the assumption of exchangeability,
which is commonplace in Bayesian statistics. Dawid (1982) 
discussed calibration (quality of match between predictive 
probabilities and the frequency of predicted events to happen) of subjectivist 
Bayesians inferences, and he suggests that badly calibrated Bayesians could do
well to adjust their future priors if this is needed to
improve calibration, even at the cost of violating coherence. 

Subjectivist Bayesianism scores well on the subjective virtues S1 and S2b.
But it is a limitation that the prior distribution exclusively formalizes
belief; context and aims of the analysis do not enter unless they have
implications about belief. In practice, an exhaustive elicitation of
beliefs is rarely feasible, and mathematical and computational convenience often plays a role in setting up
subjective priors, despite de Finetti's having famously accused frequentists 
of ``adhockeries for mathematical convenience.'' Furthermore, the assumption
of exchangeability will hardly ever precisely
match an individual's beliefs in any situation---even if there is no specific 
reason against exchangeability in a specific setup, the implicit commitment 
to stick to it whatever will be observed seems too strong---but some kind of
exchangeability assumption is required by Bayesians for the same 
reason for which frequentists need to rely on independence assumptions:  some internal replication in the model is needed to allow generalization or 
extrapolation to future observations; see Section \ref{sfreq}.  
 
Summarizing, we view much of de Finetti's criticism of frequentism as 
legitimate,
and subjectivist Bayesianism comes with a commendable honesty about 
the impact of subjective decisions and allows for flexibility accommodating
multiple perspectives. But checking and falsification of the prior is not built into the approach,
and this can get in the way of agreement between observers.
Furthermore, some problems of the frequentist approach criticized by de 
Finetti and his disciples stem from the unavoidable fact that useful 
mathematical
models idealize and simplify personal and social perspectives
on reality (see Hennig, 2010 and above), 
and the subjectivist Bayesian approach incurs 
such issues as well. 

\subsection{Objectivist Bayesianism} \label{sobjbayes}
Given the way objectivity is often advertised as a key scientific virtue (often
without specifying what exactly it means), it is not surprising that de 
Finetti's emphasis on subjectivity is not shared by all Bayesians, and that 
there have been many attempts to specify prior distributions in a more objective
way. Currently the approach of E. T. Jaynes (2003) seems to be among the most popular. As with many of his predecessors such as Jeffreys and Carnap,
Jaynes saw probability as a generalization of binary logic to uncertain 
propositions. Cox (1961) proved that given a certain list of supposedly
common-sense desiderata for
a ``plausibility'' measurement, all such measurements are equivalent, after
suitable scaling, to probability measures. This theorem is the basis of
Jaynes' objectivist Bayesianism, and the claim to objectivity comes from 
postulating that, given the same information, everybody
should come to the same conclusions regarding plausibilities: 
prior and posterior probabilities (O2c), a statement with which subjectivist 
Bayesians disagree.

In practice, this objectivist ideal seems to be
hard to achieve, and Jaynes (2003) admits that setting up objective priors 
including all information is an unsolved problem. One may wonder whether his ideal is achievable at all. For example, 
in chapter 21, he gives a full Bayesian ``solution'' to the problem of 
dealing with and identifying outliers, which assumes that prior models have
to be specified for both ``good'' and ``bad'' data (between which therefore
there has to be a proper distinction), including parameter priors for
both models, as well as a prior probability for any number of observations
to be ``bad.''  It is hard to see, and no information about this
is provided by Jaynes himself, how it can be possible to translate the
unspecific information of knowing of some outliers in many kinds of situations,
some of which are more or less related, but none identical (say) to the 
problem at hand, into precise quantitative specifications as needed for Jaynes'
approach in an objective way, all before seeing the data. 

Setting aside the difficulties or working with informally specified prior information,
even the more elementary key issue of specifying an objective prior
distribution formalizing the absence of information is riddled with
difficulties, and there are various principles for doing this which disagree in many cases (Kass and Wasserman, 
1996). Objectivity seems to be
an ambition rather than a description of what indeed can be achieved by 
setting up objectivist Bayesian priors. More modestly, therefore,
Bernardo (1979) spoke of ``reference priors,'' avoiding the
term ``objective,'' and emphasizing that it would be desirable
to have a convention for such cases (O2b), but admitting that it may not
be possible to prove any general approach for arriving at such a convention
uniquely correct or optimal in any rational sense. 

Apart from the issue of the objectivity of the specification of the prior, 
by and large the objectivist Bayesian approach has similar advantages and
disadvantages regarding our list of virtues as the subjectivist Bayesian 
approach. Particularly it comes with the same difficulties regarding the
issue of falsifiability from observations. Prior probabilities 
are connected to logical analysis of the situation rather than to betting 
rates for future observations as in de Finetti's subjectivist approach, which  makes
the connection of objectivist Bayesian prior probabilities to observations 
even weaker than in the subjectivist Bayesian approach (but
probabilistic logic has applications other than statistical data analysis, 
for which this may not be a problem).

The merit of objectivist Bayesianism is that the approach comes with a much
stronger drive to justify prior distributions in a transparent way using 
principles that are as clear and general as possible. This drive,
together with some subjectivist honesty about the fact that despite trying 
hard in the vast majority of applications the resulting prior will not deserve
the ``objectivity'' stamp and will still be subject to potential disagreement,
can potentially combine the best of both of these traditional Bayesian worlds.

\subsection{Falsificationist Bayesianism} \label{sfalbayes}
For both subjectivist and objectivist Bayesians, following 
de Finetti (1974) and Jaynes (2003), probability models including
both parameter priors and sampling models do not
model the data generating process, but rather represent plausibility or belief 
from a certain point of view. Plausibility and belief models 
can be modified by data in ways that are specified a priori, but they 
cannot be falsified by data. 

In much applied Bayesian work, on the other hand, the sampling model
is interpreted, explicitly or implicitly, 
as representing the data-generating process in a frequentist or 
similar way, and
parameter priors and posteriors are interpreted as giving information 
about what is known about the ``true'' parameter values. It has
been argued that such work does not directly run counter to the 
subjectivist or objectivist philosophy, because the ``true parameter values''
can often be interpreted as expected large sample functions given
the prior model (Bernardo and Smith, 1994), but the way in which classical
subjectivist or objectivist statistical data analysis is determined by the
untestable prior assignments is seen as unsatisfactory by many statisticians.
The suggestion of testing aspects of the prior distribution by 
observations using error statistical techniques has been around for some 
time (Box, 1980). Gelman and Shalizi (2013) incorporate this in an outline of what we refer to here as ``falsificationist Bayesianism,'' a  philosophy that openly deviates from both objectivist and
subjectivist Bayesianism, integrating Bayesian methodology with 
an interpretation of probability that can be seen as 
frequentist in a wide sense and with an error statistical approach to testing 
assumptions in a bid to improve Bayesian statistics regarding
virtue O4b.

Falsificationist Bayesianism follows the frequentist interpretation of the probabilities formalized by the
sampling model given a true parameter, so that these models can be tested using
error statistical techniques (with the limitations that such techniques have, as discussed in
Section \ref{serr}). Gelman and Shalizi argue, as some frequentists do, 
that such models are idealizations and
should not be believed to be literally true, but that the scientific process
proceeds from simplified models through test and potential falsification by
improving the models where they are found to be deficient. This reflects certain attitudes of Jaynes (2003), with the difference that Jaynes generally considered probability models as derivable from constraints of a physical system, whereas Gelman and Shalizi focus on examples in social or network science which are not governed by simple physical laws and thus where one cannot in general derive probability distributions from first principles, so that ``priors'' (in the sense that we are using the term in this paper, encompassing both the data model and the parameter model) are more clearly subjective.

A central issue for falsificationist Bayesianism is the meaning and use of the
parameter prior, which can have various interpretations, which gives 
falsificationist Bayesianism a lot of flexibility for taking into 
account multiple perspectives, contexts, and aims (S1, S2a) but may be seen as 
a problem regarding clarity and unification (O1a, O2c). Frequentists may wonder
whether a parameter prior is needed at all. Here are some
potential benefits of incorporating a parameter prior:
\begin{itemize}
\item The parameter prior may formalize relevant prior information.
\item The parameter prior may be a useful device for regularization.
\item The parameter prior may formalize deliberately extreme points of view to explore
sensitivity of the inference.
\item The parameter prior may make transparent a point of view involved in an analysis.
\item The parameter prior may facilitate a certain kind of behavior of the results
that is connected to the aims of analysis (such as penalizing complexity
or models on which it is difficult to act by giving them low prior 
weight).
\item The Bayesian procedure involving a certain parameter 
prior may have better 
error statistical properties (such as the mean squared error of point estimates derived from
the posterior) than a straightforward frequentist method, if such a method 
even exists.
\item Often finding a Bayesian parameter 
prior which emulates a frequentist/error 
statistical method helps understanding the implications of the method.
\end{itemize}
Here are some ways to interpret the parameter prior:
\begin{itemize}
\item The parameter prior may be interpreted in a frequentist way, as formalizing a more
or less idealized data generating process generating parameter values. The
``generated'' parameter values may not be directly observable, but in some
applications the idea of having, at least indirectly, a sample of several
parameter values from the parameter 
prior makes sense (``empirical Bayes''). In many 
other applications the idea is that only a single parameter from the 
parameter prior
is actually realized, which then gives rise to all the observed data. Even 
in these applications one could in principle postulate a data generating 
process behind the parameter, of which only one realization is observable, and
only indirectly. This is a rather bold idealization, but frequentists are
no strangers to such idealizations either; see Section \ref{sfreq}.
A similarly bold idealization would be to view ``all kinds of potential 
studies with the (statistically) same parameter'' as the relevant population, 
even if the studies are about different topics with different variables, 
in which case more realizations exist, but it is hard to view a specific study 
of interest as a ``random draw'' from such a population.

If parameter priors are interpreted in this sense, they can actually
be tested and falsified 
using error statistical methods; see Gelman, Meng and Stern (1996).
In situations with only one parameter realization, the power of such tests 
is low, though, and any kind of severe corroboration will be hard to achieve. 
Also, if there is only a single realization of an
idealized parameter distribution, the information in the parameter posterior
seems to rely strongly on idealization. 
\item If the quality of the inference is to be assessed by error statistical
measures,  
the parameter prior may be seen as a purely technical device. In this case, however,
the posterior distribution does not have a proper interpretation, and only 
well defined statistics with known error statistical properties
such as the mean or mode of the parameter posterior should be
interpreted.
\item Assuming that frequentist probabilities from sampling models should be 
equal to the subjectivist or objectivist epistemic probabilities if it is known 
that the sampling model is true (which Lewis, 1980, called 
``the principal principle''), the parameter prior can still be interpreted as
giving epistemic probabilities such as subjectivist betting rates, conditionally
on the sampling model to hold, even if the
sampling model is interpreted in a frequentist way. The possibility
of rejecting the 
sampling model based on the data will invalidate both coherence and Cox's 
axioms, so that the foundation for the resulting epistemic probabilities becomes
rather shaky. This does not necessarily have to stop an individual from 
interpreting and using
them as betting rates, though.  
\end{itemize}
Given such a variety of uses and meanings, it is crucial for 
applications of falsificationist Bayesianism that the choice of the 
parameter prior
is clearly explained and motivated, so transparency is central here as well 
as for the other varieties of Bayesian statistics.

Overall, falsificationist Bayesianism combines the virtue of error statistical
falsifiability with the virtues listed above as ``subjective,'' doing so via a flexibility that may be seen by some 
as problematic regarding clarity and unification. 

\section{Other philosophies}
There are important perspectives on statistics that lie outside the traditional frequentist-Bayesian divide.

In {\em machine learning}, the focus is on prediction rather than parameter estimation, thus the emphasis is on correspondence to observable reality (O4).  Computer scientists are also interested in transparency; disclosure of data, and methods with full reproducibility (O1) but are sometimes less attuned to multiple perspectives and context dependence (S1, S2).  Such attributes are necessary in practice (users have many ``knobs'' to tune in external validation, including the objective function being optimized, the division into training and test sets, and the choice of corpus to use in the evaluation)---but are typically pushed to the background.

In {\em robust statistics}, the point
is to assess stability of inferences when assumptions are violated, or to make minimal assumptions. This connects to impartiality (O3).  There is literature on classical and Bayesian robustness; in any case consideration of model violations requires awareness of multiple perspectives (S1).
Striving for robustness (against disturbances of systems, observations, assumptions) can itself be seen as a scientific virtue, although it is not normally associated with either objectivity or subjectivity.

{\em Alternative models of uncertainty} such as belief functions, imprecise probabilities or fuzzy logic aim to get around some of the limitations of probability theory (most notoriously, the difficulty of distinguishing between ``known unknowns'' and ``unknown unknowns,'' or risk and uncertainty in the terminology of Knight, 1921).  These approaches are typically framed not as subjective or objective but rather as a way to incorporate radically uncertain information into a statistical analysis.  One could say that these generalizations of probability theory aim at virtues O1c (communication of potential limitations), O2a (accounting for relevant knowledge, here regarding distinctions that are not represented in classical probability modeling)  and O4a (connection of models to observables).

{\em Exploratory data analysis} (EDA; Tukey, 1977) is all about data operations rather than models.  In that sense, EDA resembles classical statistics in its positivist focus, but with the difference that the goal is exploration rather than hypothesis testing or rigorous inference.  EDA is sensitive to multiple perspectives (S1) and context dependence (S2) in that discovery of the unexpected is always relative to what was previously expected by the researcher. Regarding transparency (O1), it could be argued that the refusal to use probability models with all their problems and particularly references to what cannot be observed (see above) contributes to clarity. However, it can also be argued that some techniques of EDA can be usefully explained in terms of probability models, e.g., as predictive checking (Gelman, 2003), but in traditional EDA such models are left incomplete or implicit, and methods that come with implicit assumptions are portrayed as assumptionless, which works against transparency.

\section{Discussion}\label{sconc}

\subsection{Implications for statistical theory and practice}

At the level of discourse, we would like to move beyond a subjective vs.\ objective shouting match.  But our goals are larger than this.  Gelman and Shalizi (2013) on the philosophy of Bayesian statistics sought not just to clear the air but also to provide philosophical and rhetorical space for Bayesians to feel free to check their models and for applied statisticians who were concerned about model fit to feel comfortable with a Bayesian approach.  In the present paper, our goals are for scientists and statisticians to achieve more of the specific positive qualities into which we decompose objectivity and subjectivity in Section \ref{svirtue}.  At the present time, we feel that concerns about objectivity are getting in the way of researchers trying out different ideas and considering different sources of inputs to their model, while an ideology of subjectivity is limiting the degree to which researchers are justifying and understanding their model.

There is a tendency for hardcore believers in objectivity to needlessly avoid the use of valuable external information in their analyses, and for subjectivists, but also for statisticians who want to make their results seem strong and uncontroversial, to leave their assumptions unexamined. We hope that our new framing of transparency, consensus, avoidance of bias, reference to observable reality, multiple perspectives, dependence on context and aims, and honesty about the researcher's position and decisions will give researchers of all stripes the impetus and, indeed, permission, to integrate different sources of information into their analyses, to state their assumptions more clearly, and to trace these assumptions backward to past data that justify them and forward to future data that can be used to validate them.

Also, we believe that the pressure to appear objective has led to confusion and even dishonesty regarding data coding and analysis decisions which cannot be motivated in supposedly objective ways; see van Loo and Romeijn (2015) for a discussion of this point in the context of psychiatric diagnosis. We prefer to encourage a culture in which it is acceptable to be open about the reasons for which decisions are made, which may at times be mathematical convenience, or the aim of the study, rather than strong theory or hard data. It should be recognized openly that the aim of statistical modeling is not always to make the model as close as possible to observer-independent reality (which always requires idealization anyway), and that some decisions are made, for example, in order to make outcomes more easily interpretable for specific target audiences.

Our key points:  (1) multiple perspectives correspond to multiple lines of reasoning, not merely to mindless and unjustified guesses; and (2) what is needed is not just a prior distribution or a tuning parameter, but a statistical approach in which these choices can be grounded, either empirically or by connecting them in a transparent way to the context and aim of the analysis.

For these reasons, {\em we do not think it at all accurate to limit Bayesian inference to ``the analysis of subjective beliefs.''} Yes, Bayesian analysis can be expressed in terms of subjective beliefs, but it can also be applied to other settings that have nothing to do with beliefs (except to the extent that all scientific inquiries are ultimately about what is believed about the world).

Similarly, {\em we would not limit classical statistical inference to ``the analysis of simple random samples.''}   Classical methods of hypothesis testing, estimation, and data reduction can be applied to all sorts of problems that do not involve random sampling. There is no need to limit the applications of these methods to a narrow set of sampling or randomization problems; rather, it is important to clarify the foundation for using the mathematical models for a larger class of problems.

\subsection{Beyond ``objective'' and ``subjective''}

The list in Section \ref{svirtue} is the core of the paper. The list may not be complete, and such a list may also be systematized in different ways. Particularly, we developed the list having particularly applied statistics in mind, and we may have missed aspects of objectivity and subjectivity that are not connected in some sense to statistics. In any case, we believe that the given list can be  helpful in practice for researchers, for justifying and explaining their choices, and for recipients of research work, for checking to what extent the listed virtues are practiced in scientific work. A key issue here is transparency, which is required for checking all the other virtues. Another key issue is that subjectivity in science is not something to be avoided at any cost, but that multiple perspectives and context dependence are actually basic conditions of scientific inquiry, which should be explicitly acknowledged and taken into account by researchers. We think that this is much more constructive than the simple objective/subjective duality.

We do not think this advice represents empty truisms of the ``mom and apple pie'' variety.  In fact, we repeatedly encounter publications in top scientific journals that fall foul of these virtues,
which indicates to us that the underlying principles are subtle and motivates this paper.  We hope that a change in names will clarify what can be done to improve statistical analyses in these two dimensions.

Instead of pointing at specific bad examples, here is a list of some issues
that can regularly be encountered in scientific publications (see, for example, our discussions in Gelman, 2015, and Gelman and Zelizer, 2015), and where we 
believe that exercising one or more of our listed virtues would improve matters:
\begin{itemize}
\item Presenting analyses that are contingent on data without explaining the exploration and selection process and without even acknowledging that it took place,
\item Justifying decisions by reference to specific literature without acknowledging that what was cited may be controversial, not applicable in the given situation, or without proper justification in the cited literature as well (or not justifying the decisions at all), 
\item Failure to reflect on whether model assumptions are reasonable in the given situation, what impact it would have if they were violated, or whether alternative models and approaches could be reasonable as well,
\item Choosing methods for the main reason that they do not require tuning or make decisions automatically and therefore seem ``objective'' without discussing whether the chosen methods can handle the data more appropriately in the given situation than alternative methods with tuning,
\item Choosing methods for the main reason that they ``do not require assumptions'' without realizing that every method is based on implicit assumptions about how to treat the data appropriately, regardless of whether these are stated in terms of statistical models,
\item Choosing Bayesian priors without justification or explanation of what they mean and imply,
\item Using nonstandard methodology without justifying the deviation from standard approaches (where they exist),
\item Using standard approaches without discussion of whether they are appropriate in the specific context.  
\end{itemize}
Most of these have to do with the unwillingness to admit to having made decisions, to justify them, and to take into account alternative possible views that may be equally reasonable.
In some sense perhaps this can be justified based on a sociological model of the scientific process in which each paper presents just one view, and then the different perspectives battle it out. But we think that this idea ignores the importance of communication and facilitating consensus for science. Scientists normally believe that each analysis aims at the truth, and if different analyses give different results, this is not because there are different conflicting truths but rather because different analysts have different aims, perspectives and access to different information. Letting the issue aside of whether it makes sense to talk of the existence of different truths or not, we see aiming at general agreement in free exchange as essential to science, and the more perspectives are taken into account, the more the scientific process is supported.

We see the listed virtues as ideals which in practice cannot generally be fully achieved in any real project. For example, tracing all assumptions to observations and making them checkable by observable data is impossible because one can always ask whether and why results from the specific observations used should generalize to other times and other situations. As mentioned in Section \ref{sfreq}, ultimately a rationale for treating different situations as ``identical and independent'' or ``exchangeable'' needs to be constructed by human thought (people may appeal to historical successes for justifying such idealizations, but this does not help much regarding specific applications). At some point---but, we hope, not too early---researchers have to resort to somewhat arbitrary choices that can be justified only by logic or convention, if that.

And it is likewise unrealistic to suppose that we can capture all the relevant perspectives on any scientific problem.  Nonetheless, we believe it is useful to set these as goals which, in contrast to the inherently opposed concepts of ``objectivity'' and ``subjectivity,'' can be approached together.

\section*{References}\sloppy

\noindent


\bibitem Alpert, M., and Raiffa, H. (1984).  A progress report on the training of probability assessors.  In {\em Judgment Under Uncertainty:  Heuristics and Biases}, ed.\ Kahneman, D., Slovic, P., and Tversky, A., 294--305.  Cambridge University Press.

\bibitem Berger, J. (2006).  The case for objective Bayesian analysis.  {\em Bayesian Analysis} {\bf 1}, 385--402.

\bibitem Bernardo, J. M. (1979). Reference posterior distributions for Bayesian inference. {\em Journal of the Royal Statistical Society B} {\bf 41}, 113--147.

\bibitem Bernardo, J. M., and Smith, A. F. M. (1994). {\em Bayesian Theory}.
Chichester: Wiley. 

\bibitem Berk, R., Brown, L., Buja, A., Zhang, K., and Zhao, L. (2013). Valid post-selection inference. {\em Annals of Statistics} {\bf 41}, 802--837.

\bibitem Box, G. E. P. (1980). Sampling and Bayes’ inference in scientific 
modelling and robustness. {\em Journal
of the Royal Statistical Society A} {\bf 143}, 383–-430.

\bibitem Box, G. E. P. (1983).  An apology for ecumenism in statistics.  In {\em Scientific Inference, Data Analysis, and Robustness}, ed.\ G. E. P. Box, T. Leonard, T., and C. F. Wu, 51--84.  New York:  Academic Press.

\bibitem Candler, J., Holder, H., Hosali, S., Payne, A. M., Tsang T., and Vizard, P. (2011). {\em  Human Rights Measurement Framework: Prototype Panels, Indicator Set and Evidence Base}. Research Report 81. Manchester: Equality and Human Rights Commission.

\bibitem Chang, H. (2012). {\em Is Water $H_2O$? Evidence, Realism and 
Pluralism}. Dordrecht: Springer.


\bibitem Cox, R. T. (1961). {\em The Algebra of Probable Inference.} Baltimore: Johns Hopkins University Press.

\bibitem Daston, L. (1992).  Objectivity and the escape from perspective.  {\em Social Studies of Science} {\bf 22}, 597--618.

\bibitem Daston, L. (1994).  How probabilities came to be objective and subjective.  {\em Historia Mathematica} {\bf 21}, 330--344.

\bibitem Daston, L., and Galison, P. (2007). {\em Objectivity}. 
New York: Zone Books.

\bibitem Davies, P. L. (2014). {\em Data Analysis and Approximate Models}.
Boca Raton, Fla.: CRC Press.

\bibitem Dawid, A. P. (1982). The well-calibrated Bayesian. {\em Journal of the
American Statistical Association} {\bf 77}, 605--610.
  
\bibitem de Finetti, B. (1974).  {\em Theory of Probability}.  New York:  Wiley.

\bibitem Desrosieres, A. (2002). {\em The Politics of Large Numbers}. Boston: Harvard University Press.

\bibitem Douglas, H. (2004). The irreducible complexity of objectivity. {\em Synthese}, {\bf 138}, 453--473.

\bibitem Douglas, H. (2009). {\em Science, Policy and the Value-Free Ideal.} University of Pittsburgh Press.


\bibitem Erev, I., Wallsten, T. S., and Budescu, D. V. (1994).  Simultaneous over- and underconfidence: The role of error in judgment processes.  {\em Psychological Review} {\bf 101}, 519--527.

\bibitem Erikson, R. S., Panagopoulos, C., and Wlezien, C. (2004).  Likely (and unlikely) voters and the assessment of campaign dynamics.  {\em Public Opinion Quarterly} {\bf 68}, 588--601.

\bibitem Everitt, B. S., Landau, S., Leese, M. and Stahl, D. (2011), 
{\em Cluster Analysis}, fifth edition. Wiley, Chichester.

\bibitem Feyerabend. P. (1978). {\em Science in a Free Society}. London: New Left Books.

\bibitem Fine, T. L. (1973). {\em Theories of Probability}. Waltham, Mass.: Academic Press.

\bibitem Fuchs, S. (1997). A sociological theory of objectivity. {\em Science Studies} {\bf 11}, 4--26.

\bibitem Gelman, A. (2003).  A Bayesian formulation of exploratory data analysis and goodness-of-fit testing. {\em International Statistical Review} {\bf 71}, 369--382.

\bibitem Gelman, A. (2008).  The folk theorem of statistical computing.  Statistical Modeling, Causal Inference, and Social Science blog, 13 May.  \url{http://andrewgelman.com/2008/05/13/the_folk_theore/}

\bibitem Gelman, A. (2013).  Whither the ``bet on sparsity principle'' in a nonsparse world? Statistical Modeling, Causal Inference, and Social Science blog, 25 Feb. \url{http://andrewgelman.com/2013/12/16/whither-the-bet-on-sparsity-principle-in-a-nonsparse-world/}

\bibitem Gelman, A. (2014a).  Basketball stats: Don't model the probability of win, model the expected score differential.  Statistical Modeling, Causal Inference, and Social Science blog, 25 Feb. \url{http://andrewgelman.com/2014/02/25/basketball-stats-dont-model-probability-win-model-expected-score-differential/}

\bibitem Gelman, A. (2014b).  How do we choose our default methods? In {\em Past, Present, and Future of Statistical Science}, ed.\ X. Lin, C. Genest, D. L. Banks, G. Molenberghs, D. W. Scott, and J. L. Wang, 293--301.  London:  Chapman and Hall.

\bibitem Gelman, A. (2014c).  President of American Association of Buggy-Whip Manufacturers takes a strong stand against internal combustion engine, argues that the so-called ``automobile'' has ``little grounding in theory'' and that ``results can vary widely based on the particular fuel that is used.''
Statistical Modeling, Causal Inference, and Social Science blog, \url{http://andrewgelman.com/2014/08/06/president-american-association-buggy-whip-manufacturers-takes-strong-stand-internal-combustion-engine-argues-called-automobile-little-grounding-theory/}

\bibitem Gelman, A. (2015). The connection between varying treatment effects and the crisis of unreplicable research: A Bayesian perspective. {\em Journal of Management} {\bf 41}, 632--643.

\bibitem Gelman, A., and Basb\o ll, T. (2013).  To throw away data: Plagiarism as a statistical crime. {\em American Scientist} {\bf 101}, 168--171. 

\bibitem Gelman, A., Bois, F. Y., and Jiang, J. (1996). Physiological pharmacokinetic analysis using population modeling and informative prior distributions. {\em Journal of the American Statistical Association} {\bf 91}, 1400--1412.

\bibitem Gelman, A., and Carlin, J. B. (2014). Beyond power calculations: Assessing Type S (sign) and Type M (magnitude) errors. {\em Perspectives on Psychological Science} {\bf 9}, 641--651. 

\bibitem Gelman, A., Carlin, J. B., Stern, H. S., Dunson, D., Vehtari, A., and Rubin, D. B. (2013).  {\em Bayesian Data Analysis}, third edition.  London:  Chapman and Hall.

\bibitem Gelman, A., and Loken, E. (2014).  The statistical crisis in science.  {\em American Scientist} {\bf 102}, 460--465.

\bibitem Gelman, A., Meng, X. L., and Stern, H. S. (1996). Posterior 
predictive assessment of model fitness
via realized discrepancies (with discussion). {\em Statistica Sinica} {\bf 6}, 
733–-807.

\bibitem Gelman, A., and O'Rourke, K. (2015).  Convincing evidence. In {\em Roles, Trust, and Reputation in Social Media Knowledge Markets}, ed.\ Sorin Matei and Elisa Bertino.  New York:  Springer.

\bibitem Gelman, A., and Shalizi, C. (2013).  Philosophy and the practice of Bayesian statistics (with discussion). {\em British Journal of Mathematical and Statistical Psychology} {\bf 66}, 8--80.

\bibitem Gelman, A., and Zelizer, A. (2015). Evidence on the deleterious impact of sustained use of polynomial regression on causal inference.  {\em Research and Politics} {\bf 2}, 1--7.

\bibitem Gillies, D. (2000).  {\em Philosophical Theories of Probability}.  London:
Routledge.

\bibitem Hacking, I. (2015). Let's not talk about objectivity. In {\em Objectivity in Science}, ed.\ F. Padovani et al. Boston Studies in the Philosophy and History of Science.   

\bibitem Hajek, A. (2009). Fifteen arguments against hypothetical frequentism. {\em Erkenntnis} {\bf 70}, 211-–235.

\bibitem Hennig, C. (2010). Mathematical models and reality: A constructivist perspective. {\em Foundations of Science} {\bf 15}, 29--48.

\bibitem Hennig, C., and Liao, T. F. (2013). How to find an appropriate clustering for mixed type variables with application to socioeconomic stratification (with discussion). {\em Journal of the Royal Statistical Science, Series C (Applied Statistics)} {\bf 62}, 309--369. 

\bibitem Hennig, C. and Lin, C.-J. (2015). Flexible parametric bootstrap for testing homogeneity against clustering and assessing the number of clusters. {\em Statistics and Computing} {\bf 25}, 821--833.

\bibitem Huber, P. J., and Ronchetti, E. M. (2009). {\em Robust Statistics}, second edition. New York: Wiley.

\bibitem Jaynes, E. T. (2003).  {\em Probability Theory:  The Logic of Science}. Cambridge University Press.

\bibitem Kahneman, D. (1999).  Objective happiness.  In {\em Well-being: Foundations of Hedonic Psychology}, 3--25. New York: Russell Sage Foundation Press.

\bibitem Kass, R. E. and Wasserman, L. (1996). The selection of prior distributions by formal rules. {\em Journal of the American Statistical Association} {\bf 91}, 1343--1370.

\bibitem Keynes, J. M. (1936).  {\em The General Theory of Employment, Interest and Money}.  London:  Macmillan.

\bibitem Knight, F. H. (1921).  {\em Risk, Uncertainty, and Profit}. Boston: Hart, Schaffner and Marx.

\bibitem Kuhn, T. S. (1962).  {\em The Structure of Scientific Revolutions}.  University of Chicago Press.

\bibitem Lewis, D. (1980). A subjectivist's guide to objective chance. In
{\em  Studies in Inductive Logic and Probability, Volume II}, ed.\ 
R. C. Jeffrey, 263-–293. Berkeley: University of California Press.

\bibitem Linstone, H. A. (1989).  Multiple perspectives:  Concept, applications, and user guidelines.  {\em Systems Practice} {\bf 2}, 307--3331.

\bibitem Little, R. J. (2012).  Calibrated Bayes, an alternative inferential paradigm for official statistics.  {\em Journal of Official Statistics} {\bf 28}, 309--334.

\bibitem MacKinnon, C. (1987). {\em Feminism Unmodified}. Boston: Harvard University Press.

\bibitem Maturana, H. R. (1988). Reality: The search for objectivity or the
quest for a compelling argument. {\em Irish Journal of Psychology} {\bf 9}, 
25--82.

\bibitem Mayo, D. G. (1996). {\em Error and the Growth of Experimental Knowledge}. University of
Chicago Press.

\bibitem Mayo, D. G. (2014).  Objective/subjective, dirty hands, and all that.  Error Statistics Philosophy blog, 16 Jan. \url{http://errorstatistics.com/2014/01/16/objectivesubjective-dirty-}\\
\url{hands-and-all-that-gelmanwasserman-blogolog/}

\bibitem Mayo, D. G. and Spanos, A. (2010).  Introduction and background: The error-statistical philosophy. In {\em Error and Inference}, ed. Mayo, D. G. and Spanos, A., 15--27. Cambridge University Press. 

\bibitem Megill, A. (1994). Introduction: Four senses of objectivity. In {\em Rethinking Objectivity}, ed. A. Megill, 1--20. Durham, N.C.: Duke University Press.

\bibitem Merry, S. E. (2011). Measuring the world:  Indicators, human rights, and global governance. {\em Current Anthropology} {\bf 52} (S3), S83--S95.


\bibitem Pearson, K. (1911). {\em The Grammar of Science}. 2007 edition. 
New York: Cosimo.

\bibitem Pollster.com (2004). Should pollsters weight by party identification? \url{http://www.pollster.com/faq/should_pollsters_weight_by_par.php}

\bibitem Porter, T. M. (1996).  {\em Trust in Numbers: The Pursuit of Objectivity in Science and Public Life}. Princeton University Press. 

\bibitem Reichenbach, H. (1938). On probability and induction, {\em Philosophy of Science}, {\bf 5}, 21-–45.

\bibitem Reiss, J., and Sprenger, J. (2014). Scientific objectivity. In {\em Stanford Encyclopedia of Philosophy} (Fall 2014 Edition), ed.\ E. N. Zalta, \url{http://plato.stanford.edu/archives/fall2014/entries/scientific-objectivity/}.

\bibitem Rubin, D. B. (1978).  Bayesian inference for causal effects:  The role of randomization.  {\em Annals of Statistics} {\bf 6}, 34--58.

\bibitem Rubin, D. B. (1984).  Bayesianly justifiable and relevant frequency calculations for the applied statistician.  {\em Annals of Statistics} {\bf 12}, 1151--1172.

\bibitem Saari, C. (2005).  The contribution of relational theory to social work practice.  {\em Smith College Studies in Social Work} {\bf 75}, 3--14.

\bibitem Sheiner, L. B. (1984).  The population approach to pharmacokinetic data analysis:  Rationale and standard data analysis methods.  {\em Drug Metabolism Reviews} {\bf 15}, 153--171.

\bibitem Silberzahn, R., et al.\ (2015).  Crowdsourcing data analysis: Do soccer referees give more red cards to dark skin toned players?  Center for Open Science, \url{https://osf.io/j5v8f/}

\bibitem Simmons, J., Nelson, L., and Simonsohn, U. (2011). False-positive psychology: Undisclosed flexibility in data collection and analysis allow presenting anything as significant. {\em Psychological Science} {\bf 22}, 1359--1366.

\bibitem Tibshirani, R. J. (2014).  In praise of sparsity and convexity. In {\em Past, Present, and Future of Statistical Science}, ed.\ X. Lin, C. Genest, D. L. Banks, G. Molenberghs, D. W. Scott, and J. L. Wang, 505--513.  London.:  Chapman and Hall.

\bibitem van Fraassen, B. (1980). {\em The Scientific Image}. Oxford University Press.

\bibitem van Loo, H. M., and Romeijn, J. W. (2015).  Psychiatric comorbidity: Fact or artifact?  {\em Theoretical Medicine and Bioethics} {\bf 36}, 41--60.

\bibitem von Glasersfeld, E. (1995). {\em Radical Constructivism: A Way of 
Knowing and Learning}. London: Falmer Press.

\bibitem von Mises, R. (1957). {\em Probability, Statistics and Truth}, second revised English edition. New York: Dover.
 
\bibitem Wang, W., Rothschild, D., Goel, S., and Gelman, A. (2015).  Forecasting elections with non-representative polls. {\em International Journal of Forecasting} {\bf 31}, 980--991.

\bibitem Weinberger, D. (2009).  Transparency is the new objectivity.  Everything is Miscellaneous blog, 19 Jul. \url{http://www.everythingismiscellaneous.com/2009/07/19/transparency-is-the-}\\
\url{new-objectivity/}

\bibitem Yong, E. (2012).  Nobel laureate challenges psychologists to clean up their act.  {\em Nature News}, 3 Oct.  \url{http://www.nature.com/news/nobel-laureate-challenges-psychologists-to-clean-}\\
\url{up-their-act-1.11535}

\bibitem Zabell, S. L. (2011).  The subjective and the objective.  In {\em Philosophy of Statistics}, ed.\ P. S. Bandyopadhyay and M. R. Foster.  Amsterdam:  Elsevier.

\end{document}